\documentclass[a4paper,fleqn,usenatbib]{mnras}

\usepackage{newtxtext,newtxmath}

\usepackage[T1]{fontenc}
\usepackage{ae,aecompl}
\usepackage{xcolor}

\usepackage{graphicx}	
\usepackage{amsmath}	

\title[Search for the origin of Galactic CRs with the PTS]{Search for the Galactic accelerators of Cosmic-Rays up to the Knee with the Pevatron Test Statistic}

\author[Ang\"uner et al.]{
Ekrem O\u{g}uzhan Ang\"uner,$^{1}$\thanks{oguzhan.anguner@tubitak.gov.tr (E.O.~Ang\"uner)}
Gerrit Spengler,$^{2}$\thanks{spengler@physik.hu-berlin.de (G.~Spengler)}
Elena Amato,$^{3,4}$\thanks{elena.amato@inaf.it (E.~Amato)}
Sabrina Casanova$^{5}$\thanks{sabrina.casanova@ifj.edu.pl (S.~Casanova)}
\\
$^{1}$TÜBİTAK Research Institute for Fundamental Sciences, 41470 Gebze, Turkey\\
$^{2}$Institut f\"ur Physik, Humboldt-Universit\"at zu Berlin, Newtonstr. 15, 12489 Berlin, Germany\\
$^{3}$INAF - Osservatorio Astrofisico di Arcetri, Largo E. Fermi, 5, 50125, Firenze, Italy\\
$^{4}$ Dipartimento di Fisica e Astronomia, Universit\`a di Firenze, Via Sansone 1, I-50019 Sesto Fiorentino (FI), Italy\\
$^{5}$IFJ Institute of Nuclear Physics, Radzikowskiego 152, 31-342, Krakow, Poland\\
}

\date{Accepted 27.05.2023. Received 20.04.2023}

\pubyear{2023}

\begin{document}
\label{firstpage}
\pagerange{\pageref{firstpage}--\pageref{lastpage}}
\maketitle

\begin{abstract}
The Pevatron~Test~Statistic~(PTS) is applied to data from $\gamma$-ray observatories to test for the origin of Cosmic~Rays~(CRs) at energies around the knee of the CR spectrum. Several sources are analyzed within hadronic emission models. Previously derived results for RX~J1713.7$-$3946, Vela~Jr., and HESS~J1745$-$290 are confirmed to demonstrate the concept, reliability, and advantages of the PTS. It is excluded with a significance more than $5\sigma$ that the sources RX~J1713.7$-$3946 and Vela~Jr.~are Pevatrons, while strong indications exceeding $4\sigma$ are found for excluding HESS~J1745$-$290 as a Pevatron. The importance to resolve source confusion with high angular resolution observations for Pevatrons searches is demonstrated using PTS for the region containing the SNR~G106.3+2.7 and the Boomerang nebula. No statistically significant conclusion with respect to Pevatron associations could be drawn from this region, for the diffuse $\gamma$-ray emission around the Galactic~Center, and the unidentified $\gamma$-ray sources LHAASO~J2108$+$5157, HESS~J1702$-$420A and MGRO~J1908$+$06. Assuming the entire $\gamma$-ray emission from MGRO~J1908+06 and the tail region of SNR~G106.3+2.7 is hadronic, a statistical indication exceeding $3\sigma$ is found for the underlying proton spectrum to extend beyond 350–400~TeV as a power-law. This result can indicate that these sources are proton and helium Pevatrons, in which the accelerated particles contribute to the knee of proton and helium spectra observed at Earth.
\end{abstract}




\section{Introduction}
\label{sec_intro}
The Cosmic~Rays~(CRs) that enter the atmosphere of the Earth have now been investigated for more than a century after their first detection \citep{hess_discovery}, for which the year 1936 Nobel prize was awarded. As, for example, reviewed in \cite{elena_rev, elena_rev2, blasi_rev}, the flux of CRs detected on Earth is dominated by protons, with helium being the second most abundant nucleus. The energy spectrum above $\sim 30$~GeV up to the so-called "knee" is very well approximated by a power-law with spectral index $-2.7$, although significant deviations from this simple model have been recently detected. The "knee" is a prominent feature seen in the CR energy spectrum at $\sim$3~PeV energies, where the spectral index steepens significantly to $\sim-3.0$. Although some recent evidence exists that the knee might be below $1$~PeV when only the combination of protons and helium nuclei is considered \citep{knee_below_1pev}, it is clear that, at least for heavier elements, the spectral steepening occurs at energies well above $1$~PeV~\citep{z_dependence_knee}. The origin of the knee is debated ever since its first discovery \citep{knee_discovery}, with two interpretations being particularly popular. As reviewed in \cite{blumer_rev}, the first model identifies the knee energy with the maximum achievable energy of Galactic particle accelerators, while the second model proposes a connection between the knee and the maximum energy for which electrically charged particles are magnetically confined within the Galaxy. In addition to the origin of the knee, it remains to this date an open question whether the sites where CRs are accelerated up to or beyond the energy of knee are within the Galaxy. 

A Pevatron is in the following defined to be a source of CRs at energies around the knee of the CR spectrum. The localization of Pevatrons within the Galaxy would therefore positively decide the question of whether CRs are accelerated within the Galaxy up to the knee of the CR spectrum. From a theoretical side, multiple plausible astrophysical objects, with young remnants of Supernovae \citep{baade_zwicky, ginzburg} above all, were proposed, as reviewed for example in \cite{pierre_rev}. However, no Galactic source showing firm evidence of hadronic acceleration to PeV energies and beyond has been identified to this date. The Pevatron~Test~Statistic~(PTS), which offers a new approach to detect spectral signatures of Pevatrons, was recently introduced in \cite{cta_pevatron} to estimate the sensitivity of the planned Cherenkov~Telescope~Array~(CTA) to Pevatron sources. 

In this paper, the PTS \citep{cta_pevatron} is applied for the first time to publicly available spectral data from different $\gamma$-ray observatories. The aim is to test whether the sources of the respective $\gamma$-rays are Pevatrons. The paper is structured as follows. Motivation for the stated definition of a Pevatron is briefly discussed in Sec.~\ref{sec_pevatron_def}. The principle for the identification of Pevatrons by means of $\gamma$-ray spectra is discussed in Sec.~\ref{sec_spectral_sigs}, together with a brief assessment of the advantages of the PTS compared to other currently employed methods for the detection of Pevatrons. The calculation and interpretation of the PTS for public data from a selection of $\gamma$-ray sources is discussed in Sec.~\ref{sec_data_analyis_all}. The PTS profiles of Pevatron candidate sources are provided and discussed in Sect.~\ref{pts_profile}. Finally, the conclusions are summarized in Sec.~\ref{sec_conclude}.

\section{What is a Pevatron?}
\label{sec_pevatron_def}

Two different definitions for a Pevatron are currently used in the literature and discussed in \cite{alison}. A Pevatron is defined in both cases as an astrophysical source in which individual particles are accelerated to energies beyond $1$~PeV. However, in one case the name is reserved for hadronic accelerators while, in the other case, it is additionally used to denote leptonic accelerators. In the following, a Pevatron is defined to be a source of CRs at energies around the knee of the CR spectrum. This definition is briefly motivated and discussed in the following.

The Tevatron, built at Fermilab \citep{tevatron}, was able to accelerate particles to TeV energies. This was indicated in the name 'Tevatron', which is a contraction of the metric prefix for the maximum achievable energy, and the Greek word 'tron' for 'tool'. Following this scheme, a Pevatron is literally a tool to accelerate particles to at least an energy of $1$~PeV. The application of the term in astrophysics faces the problem that the astrophysical accelerators are not purposely used tools, but they are themselves the objects of study whose physical principles are under investigation. Instead, in recent astrophysical practice regarding Pevatrons, the maximum achievable particle energy of the accelerator is often considered to be eponymous. In this approach, an astrophysical Pevatron is an accelerator with a maximum energy of at least $1$~PeV. This definition applies to accelerators of hadrons as well as electron accelerators such as the Crab nebula, which has been known for at least a decade to host PeV leptons (see e.g. \cite{AmatoOlmi21} for a review) and from which photons with energies above $1$~PeV were recently detected \citep{crab_pev}.

From a historical perspective, however, the term Pevatron is introduced in astrophysics to denote the putative sources of CRs at the knee of the CR spectrum. The focus here is not primarily on the maximum energy of the accelerator, but the introduction of the term Pevatron is justified by the presence of the knee in the CR spectrum which suggests a new physical effect on the scale of the Galaxy, as discussed in Sec.~\ref{sec_intro}. As a consequence, the maximum energy of $1$~PeV is not considered as the primary property of a Pevatron. Instead, a Pevatron is in the following defined to be a source of CRs with energies around the knee of the CR spectrum. The search for Pevatrons is then connected with the broader quest for the origin of CRs. In general, features in the CR spectrum might be related either to their acceleration or propagation \citep{elena_rev}. As mentioned before, features exist in the CR spectrum detected on Earth also at energies lower than the knee, most notably a hardening observed in all nuclear species at around 300~GeV \citep{AMS02protons,AMS02PrimNuclei,CREAM10}. However, as testified by the differences between the spectra of primary and secondary nuclei \citep{AMS02Secondaries}, these must be related to the physics of propagation in the Galaxy (see e.g. \citet{elena_rev, elena_sabrina} for a detailed discussion). 

The knee is then the lowest energy feature that might be directly related to the properties of CR accelerators. For a long time, the general consensus has been in favour of the identification of this feature with the maximum energy achievable by CR protons in Galactic sources. The steepening observed at around 1~PeV would result from the superposition of the cutoffs of different CR elements, with heavier, less abundant elements reaching higher maximum energies thanks to the rigidity dependence of the acceleration mechanism. The above mentioned recent evidence for a knee at slightly lower energy than 1~PeV, when only protons and He nuclei are considered, does not change the picture much: the best estimate for the p+He knee is $E_{\rm knee}(p+He)= 700$ TeV \citep{knee_below_1pev}, less than a factor two different and still compatible with 1 PeV within the uncertainties. The conclusion is that, if the knee really is a signature of the maximum rigidity that Galactic accelerators can provide, the primary CR sources in the Galaxy must be able to accelerate particles at least up to 1 PeV. This definition has three important implications:
\begin{itemize}
    \item[1:] A Pevatron must accelerate hadrons.
    \item[2:] Because the energy of the knee is, at least for heavy elements, well above $1$~PeV, the maximum energy of a Pevatron must be much larger than $1$~PeV.
    \item[3:] It must be possible to explain the steepening of the CR spectrum at the knee by a combination of intrinsic properties of the Pevatron and propagation effects.
\end{itemize}
\section{Search for spectral signatures of Pevatrons with Gamma-ray observatories}
\label{sec_spectral_sigs}
Deflection of charged particles by Galactic magnetic fields prevents direct localization of Pevatrons through CR measurements on Earth. Instead, indirect fingerprints of the presence of Pevatron activities must be searched for. Such fingerprints emerge from pp-interactions, namely the interactions of hadrons accelerated in a Pevatron with target material. The latter can easily be traced and determined from infra-red, sub-millimeter and radio observations \citep{roman_duval_radio} and it is an astronomical multi-messenger problem to detect the electrically neutral secondary particles, more concretely neutrinos and $\gamma$-rays, which are created in the pp-interactions.

In the following, signatures of Pevatrons are searched for with a statistical test based on a hadronic model which reproduces the observed $\gamma$-ray emission, as discussed below in Sec.~\ref{gamma_model_sec}, and spectral data acquired from different $\gamma$-ray observations of various sources. The advantages of the method over other currently used search methods are discussed in Sec.~\ref{sec_pts}.

\subsection{Spectral gamma-ray signatures of Pevatrons}
\label{gamma_model_sec}

The differential energy distribution of accelerated hadrons, n($E_\mathrm{p}$), is in the following assumed to follow a simple power-law with spectral index $\Gamma_\mathrm{P}$ and an exponential cutoff at an energy $E_\mathrm{cut,\,p}$, with sharpness described by the parameter $\beta$:
\begin{equation}
n(E_\mathrm{p}) \sim E_\mathrm{p}^{-\Gamma_\mathrm{P}}\;\exp\left(-\left(\frac{E_\mathrm{p}}{E_\mathrm{cut,\,p}}\right)^{\beta}\right)\,\mathrm{.}
\label{eq_model}
\end{equation}

The exact shape of the cutoff, namely the value of $\beta$, depends in principle on what limits the acceleration. Assuming that the main mechanism responsible for CR acceleration is Diffusive Shock Acceleration (DSA), as is the case for the most commonly invoked potential sources, such as SNRs \citep{pierre_rev} or Young Massive Star Clusters \citep{AharonianSFR,MorlinoWind,BykovSFR}, the most stringent limitation is usually provided by the size of the accelerator compared to the diffusion distance of the highest energy particles. This translates into the condition $D(E_{\rm max})$=$v_s$~L, where $D$ is the diffusion coefficient, $v_s$ is the shock velocity and $L$ is the size of the accelerator (i.e.~the radius of the SNR or of the wind termination shock in the case of a star cluster). Writing the diffusion coefficient as $D(E)=D_0 E^\delta$, it is possible to show that the particle spectrum turns out to be the one in Eq.~\ref{eq_model} with $\beta=\delta$ \citep{expcutoff_ref}. In particular, an exponential cutoff is found for Bohm diffusion ($\delta=1$), while sub-exponential cutoffs result from other diffusion models commonly adopted in astrophysics, such as Kolmogorov's ($\delta=1/3$) or Kraichnan's ($\delta=1/2$). 

Equation~\ref{eq_model} still provides a good description of the particle spectrum in scenarios that connect the maximum particle energy to magnetic field growth (see e.g.\citet{Schure13,Cristofari20}). Current theories of efficient acceleration at shocks assume that the magnetic turbulence responsible for particle diffusion is self-generated by the particles being accelerated. As far as SNRs are concerned, in particular, the most common view is that achieving energies close to the knee is only made possible by the so-called non-resonant streaming instability \citep{Bell04}, induced by the particles at the instantaneous maximum energy leaving the accelerator. In these scenarios, the maximum energy is connected to the magnetic field growth, rather than limited by the system size (see e.g.\citet{Schure13,Cristofari20}) and the instantaneous spectrum at the shock is usually assumed to be cut very sharply at $E_{\rm max}$, which would reflect the case of super-exponential cutoffs ($\beta>1$) in Eq.~\ref{eq_model}. 

\begin{figure}
\centering
\includegraphics[width=9.4cm]{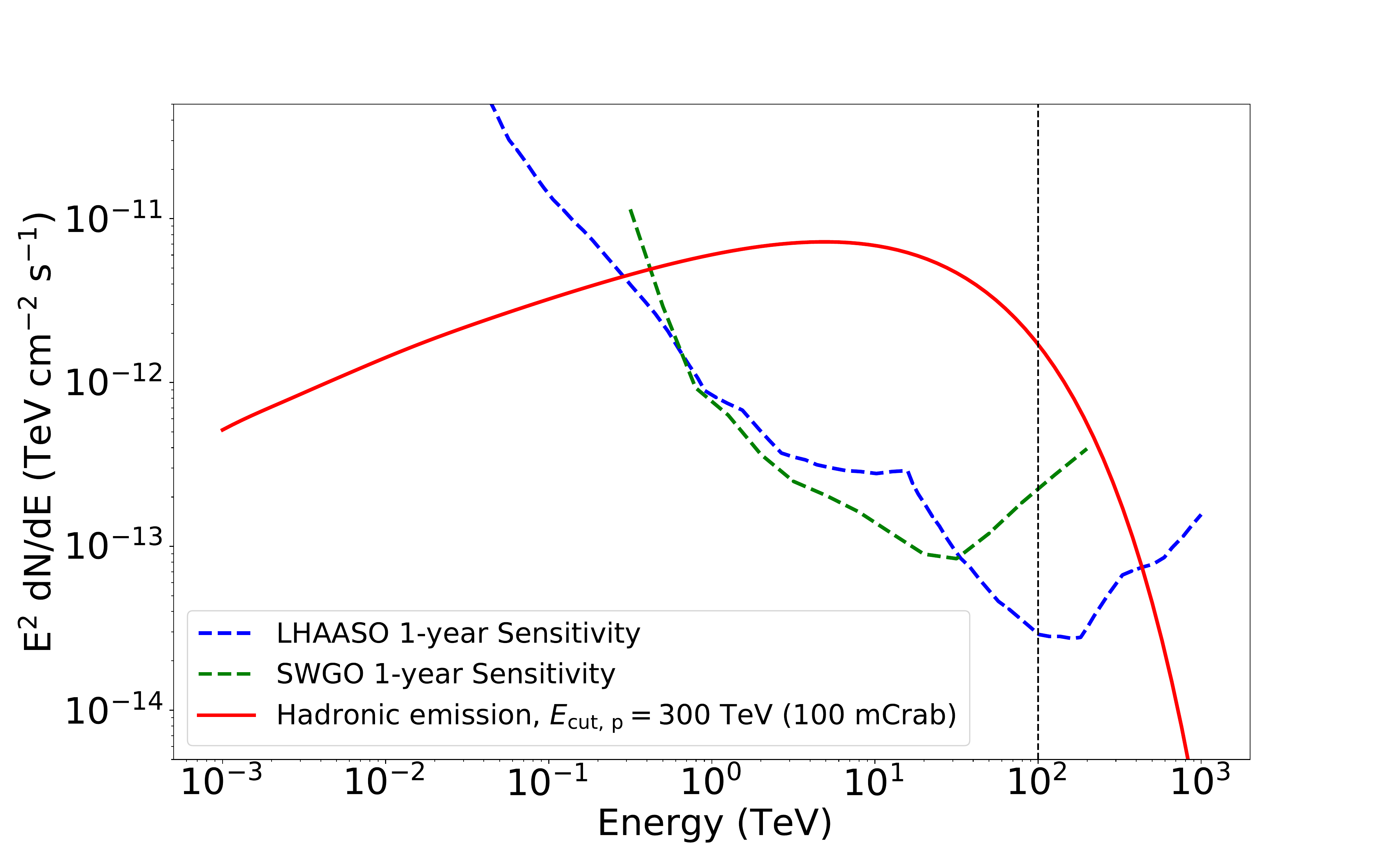}
\caption{An example of the predicted $\gamma$-ray spectrum which results from interactions of protons with gas is shown in red. The energy spectrum of the protons is assumed to be given by Eq.~\ref{eq_model}, with parameters $\Phi_0=100$~mCrab (at 1 TeV), $\Gamma_\mathrm{P}=1.7$ and $E_\mathrm{cut,\,p}=300$~TeV, i.e. the hadron accelerator is not a Pevatron. Shown for comparison are the 1$-$year differential flux sensitivity of LHAASO (blue) and the 1-year reference configuration sensitivity of SWGO \citep{swgo_white} (green) taken from https://github.com/harmscho/SGSOSensitivity. The vertical black line indicates the threshold energy of $100$~TeV above which a flux detection is often interpreted as an indication for the location of a Pevatron.}
\label{fig_sens}
\end{figure}

The $\gamma$-ray emission $\Phi_\gamma(E)$ created in interactions of accelerated hadrons with ambient gas is calculated with {\tt naima} package \citep{naima}, assuming the pp-cross section derived in \cite{naima_pi0}. In practice, a normalization of Eq.~\ref{eq_model} is calculated for a predicted $\gamma$-ray spectrum given a spectral index $\Gamma_\mathrm{P}$ and an energy cutoff $E_\mathrm{cut,\, p}$. Instead of a direct normalization of the proton spectrum, the predicted $\gamma$-ray flux $\Phi_0$ at an energy of $E_\gamma=1$~TeV is used as a normalization parameter for the hadron spectrum $n(E_\mathrm{p})=n(E_\mathrm{p}|E_\mathrm{cut,\,p}, \boldsymbol{\theta})$ where $\boldsymbol{\theta}=(\Gamma_\mathrm{P},\Phi_0)$. This convention simplifies the interpretation of the predicted flux in the context of $\gamma$-ray detectors.

As an example, Fig.~\ref{fig_sens} shows the predicted $\gamma$-ray spectrum resulting from a proton spectrum with spectral index of $\Gamma_\mathrm{P}=1.7$, an energy cutoff of $E_\mathrm{cut,\,p}=300$~TeV and differential flux of $\Phi_0=100$~mCrab\footnote{Throughout the paper, Crab unit is assumed as the differential Crab flux at 1 TeV of 3.84~$\times$ 10$^{-11}$ cm$^{-2}$ s$^{-1}$ TeV$^{-1}$, taken from Table 6 of \cite{hess_crab}} at $1$~TeV. As discussed in more detail in \cite{aharonian_gamma} and \cite{celli_2020}, the resulting $\gamma$-ray spectrum is itself well described by a power-law with index $\Gamma_\mathrm{\gamma}$ = $\sim$$\Gamma_\mathrm{P}-0.15$ and sub-exponential cutoff. 

\subsection{The PTS and other criteria for the Pevatron detection}
\label{sec_pts}
Given a set of observational data $\mathrm{D}$, the best fit parameters $E^*_\mathrm{cut,\,p}$ and $\boldsymbol{\theta}^*$ for the hadronic emission model discussed above in Sec.~\ref{gamma_model_sec} can be determined through the maximization of a likelihood function $L(E_\mathrm{cut,\,p},\,\boldsymbol{\theta}|\mathrm{D})$. In the following, only flux data $\Phi(E_i)$ with errors $\sigma(E_i)$ in energy bins $E_i$ are analyzed, and the likelihood function is given by 

\begin{equation}
L(E_\mathrm{cut,\,p},\,\boldsymbol{\theta}|D)=-2\sum_i \left(\frac{\Phi_\gamma(E_i|E_\mathrm{cut,\,p},\,\boldsymbol{\theta})-\Phi(E_i)}{\sigma(E_i)}\right)^2\,\mathrm{.}
\end{equation}

The PTS
\begin{equation}
\label{eq_PTS}
\mathrm{PTS}=-2\ln\frac{\hat
L(E_\mathrm{cut,\,p}=1\,\mathrm{PeV},\boldsymbol{\theta}|D)}{\hat L(E_\mathrm{cut,\,p}, \boldsymbol{\theta}|D)}\,\mathrm{,}
\end{equation}
is introduced in \cite{cta_pevatron} as a likelihood ratio test for the deviation of the energy cutoff $E_\mathrm{cut,\,p}$ in Eq.~\ref{eq_model} from $1$~PeV. $\hat{L}(E_\mathrm{cut,\,p},\boldsymbol{\theta}|D)$ is the maximum of the likelihood over all values for $E_\mathrm{cut,\,p}$ and $\boldsymbol\theta$, including negative values for $E_\mathrm{cut,\,p}$, and $\hat{L}(E_\mathrm{cut,\,p}$=1\,$\mathrm{PeV}$,$\boldsymbol{\theta}|D)$ is the maximum likelihood when the cutoff energy is fixed to the Pevatron threshold of $1$~PeV. The statistical significance of the PTS is calculated as 

\begin{equation}
\label{eq_S_PTS}
S_\mathrm{PTS}=\mathrm{sign}(E_\mathrm{cut,\,p}^*-1\,\mathrm{PeV})\sqrt{\mathrm{PTS}}\ .
\end{equation}
For $S_\mathrm{PTS}<-5$, the association of a $\gamma$-ray source with a Pevatron can be excluded with a CL corresponding to at least $5\sigma$. If, on the other hand,  $S_\mathrm{PTS} \geq 5$, a Pevatron detection can be claimed with a CL corresponding to at least $5\sigma$ under the assumption that the detected $\gamma$-ray emission is generated in interactions of hadrons with target nuclei.~In other words, $S_\mathrm{PTS}>5$ ensures with a CL corresponding to at least $5\sigma$ that the underlying hadron spectrum goes well beyond 1 PeV as power-law without showing any signs of a spectral cutoff, and consequently, such a source contributes to the CR spectrum at energies above 1 PeV. For $|S_\mathrm{PTS}|<5$, the data are insufficient to decide whether or not the $\gamma$-ray source is associated with a Pevatron, and typically more data must then be acquired to make a decision based on the PTS possible. In the hypothetical case where the true cutoff energy $E_\mathrm{cut,\,p}$ is equal to the threshold energy of $1$~PeV, the PTS is by definition insensitive given finite data. In practice, the PTS can only detect a Pevatron when the true cutoff energy is much larger than $1$~PeV. This reflects the Pevatron definition discussed in Sec.~\ref{sec_pevatron_def} according to which a Pevatron must accelerate hadrons to energies well above $1$~PeV. More information on the interpretation of $S_\mathrm{PTS}$ and the connection between $S_\mathrm{PTS}$ and the PTS can be found in \cite{cta_pevatron}.

Two alternative methods, the detection significance of the $\gamma$-ray emission above $100$~TeV and the 95$\%$ CL lower limit of the hadronic energy cutoff, are currently used in the literature to claim evidence for a Pevatron detection. The claim for the presence of a Pevatron based on a lower limit on the energy cutoff inferred to be larger than $1$~PeV faces the problem that the confidence level, typically $95\%$ or less\footnote{The z-score of $95\%$ C.L. is $\sim$1.96.}, is much smaller than the confidence level corresponding to $5\sigma$, which is typically requested for a detection. On the other hand, detection of a significant (i.e. 5$\sigma$) cutoff in the hadronic energy spectrum well below 1 PeV (i.e. $E_\mathrm{cut,\,p}\ll1$~PeV) can serve as strong evidence against a potential association between a $\gamma$-ray source and a Pevatron. This asymmetry between the confidence level used for exclusion and confirmation of an association between a $\gamma$-ray source and a Pevatron is unsatisfactory, in particular when one deals with such an important claim as the detection of the sources of the highest energy CRs in the Galaxy, which certainly deserves to be made with a high confidence level. Similarly, the association between a $\gamma$-ray source with significant ($>$ 5$\sigma$) emission at energies greater than $100$~TeV and a Pevatron is problematic.~Figure~\ref{fig_sens} shows a $\gamma$-ray spectrum predicted for pp-interactions given a true hadronic cutoff energy of $E_\mathrm{cut,\,p}=300$~TeV, i.e. for a hadronic accelerator which is not a Pevatron, together with the sensitivities of the Large High Altitude Air Shower Observatory (LHAASO) and the planned SWGO. It is obvious that both, SWGO and LHAASO, would be able to detect significant $\gamma$-ray emission from this simulated source above energies of $100$~TeV, although this source is not associated with a Pevatron. The problem with this method is that a Pevatron is identified with cumulative excess events above 100 TeV and independent of the spectral shape, which does not guarantee that the cut-off energy is well above 1 PeV.\\
The PTS method avoids both problems: confirmation and rebuttal of the association between a $\gamma$-ray source and a Pevatron are assessed with the same confidence level and the spectral shape is employed to ensure that the hadron energy cutoff is well above $1$~PeV when detection is claimed. Figure 8 of \cite{cta_pevatron} shows the relation between PTS and 95$\%$ CL lower limit on the proton spectral cutoff, together with the significance of E$>$100 TeV detection obtained from simulations of synthetic Pevatron sources. It was shown that these properties are strongly correlated and requirements for both of these alternative methods are well satisfied when the condition $S_\mathrm{PTS} \geq 5$ is satisfied. 

\section{Application to data}
\label{sec_data_analyis_all}
The PTS is in the following calculated and interpreted for selected $\gamma$-ray sources based on public spectral data. As a first test, the PTS is calculated for three sources that are not considered to be Pevatrons, and results already established are confirmed with the new PTS concept. The discussion starts with the two shell-type SNRs, Vela~Jr. and RX~J1713.7$-$3946. Afterwards, HESS~J1745$-$290, which is spatially coincident with the compact radio source Sgr~A* at the center of the Galaxy, is discussed. For these three sources, results that were derived previously by other means and proving the non-Pevatron nature, are confirmed with the PTS.

In a second step, it is shown that the PTS cannot decide whether the diffuse $\gamma$-ray emission from the vicinity of the Galactic Center (GC) is emitted by interactions of hadrons which are accelerated in a Pevatron.~Both Pevatron \citep{hess_gc_pevatron, veritas_gc} and non-Pevatron \citep{magic_diffuse_gc} conclusions were previously drawn for the diffuse $\gamma$-ray emission based on the derived lower limit on the hadronic cutoff energy. Together with the previous examples, this discussion shows the ability of the PTS to either decide whether a $\gamma$-ray source is a Pevatron at a given significance level or to quantify that a decision is impossible based on the available data, with the same unified criterion.

The PTS is applied to the recently detected   ultra-high-energy (UHE, E$>$100~TeV) Pevatron candidate $\gamma$-ray sources of LHAASO~J2226$+$6057, MGRO~J1908$+$06, LHAASO~J2108$+$5157, and HESS~J1702$-$420A. The potential of using high angular resolution observations to resolve source confusion and locate Pevatrons is demonstrated and explored based on the PTS analysis of the LHAASO~J2226$+$6057 region. The joint spectral analysis of the LHAASO~J2226$+$6057 region using PTS results in a significant rejection of the Pevatron hypothesis, when source confusion cannot be 
resolved. However, by using spectral data from high angular resolution observations to address source confusion, a sub-component of this region emerges as one of the best Pevatron candidate. In any case, additional spectral data are needed for these sources to decide whether they are associated with hadronic Pevatrons that can explain the 3 PeV knee feature.

Finally, the PTS profiles of the Pevatron candidates are extracted. It is argued that the proton spectra underlying the observed $\gamma$-ray emission from MGRO~J1908$+$06 and the tail region of SNR~G106.3$+$2.7 can reach a marginal $S_\mathrm{PTS}$ significance level of 3$\sigma$ at energies around 350--400 TeV (and 5$\sigma$ at 150--200 TeV). Assuming that the knee of proton (and helium) spectra observed from the Earth is below 1~PeV (i.e.~$\sim$700~TeV \citep{knee_below_1pev}), then the fact that these sources have reached marginal $S_\mathrm{PTS}$ levels suggests that they could be responsible for contributing to the knee of the proton spectra. Therefore, it is possible that these sources are proton Pevatrons, although the evidence for this contribution is only marginally significant.

\subsection{Data analysis}
\label{data_analysis}
In the following sections, public spectral $\gamma$-ray flux data from observations of different sources are analyzed. For each source, a flux dataset contains estimates of the differential $\gamma$-ray flux, $dN/dE$, at different energies. Flux measurements inferred from data acquired with different instruments are analyzed jointly in the framework of {\tt gammapy} \citep{gammapy_v018}. Where asymmetric statistical errors, $[\sigma_-,\sigma_+]$, are reported for a $\gamma$-ray flux point, a conservative symmetric statistical error $\sigma_\mathrm{stat}=\mathrm{max}\{\sigma_-,\,\sigma_+\}$ is used. Additionally, a systematic error $\sigma_\mathrm{sys}$ on each differential flux point is considered. The systematic error is assumed to scale proportionally to the estimated flux, i.e. $\sigma_\mathrm{sys}=\xi\,dN/dE$, where $\xi$ can be considered as the minimal relative error that is considered for each flux point. In the following, all conclusions are based on a conservative relative flux error of $\xi\geq 20\%$. Analyses with $\xi<20\%$ are only discussed to the purpose of illustrating the dependence of the analysis on the assumed value of systematics error, $\xi$.
The final conservative error on each differential flux point is calculated as $\sigma=\mathrm{max}\{\sigma_\mathrm{sys},\,\sigma_\mathrm{stat}\}$.

For each considered $\gamma$-ray source, the respective flux dataset is fitted to a hadronic $\gamma$-ray emission model as described in Sec.~\ref{gamma_model_sec}, and the best-fit parameters are derived from $\chi^2$-minimization. Lower limits on the hadronic cutoff energy $E_\mathrm{cut,\,p}$ and the significance $S_\mathrm{PTS}$ of the PTS are derived as detailed in \cite{cta_pevatron} with the {\tt ecpli} package~\citep{ecpli}. The reported p-values are derived from a $\chi^2$ test of the best fit model against the spectral data, with the error $\sigma$ defined as above.

The $\gamma$-ray flux, $\phi_\mathrm{true}$, emitted by a source is attenuated due to the effect of pair creation on interstellar radiation fields, i.e. the process  $\gamma\gamma\rightarrow e^+e^-$ also known as $\gamma\gamma$-absorption. Following \cite{absorption_vernetto}, it is assumed that the probability $1-P$ for a $\gamma$-ray to be absorbed due to pair creation within the Galaxy is smaller than $10\%$ for $\gamma$-ray energies below $100$~TeV. The relative correction to the observed flux due to $\gamma\gamma$-absorption, $(\phi_\mathrm{true}-\phi_\mathrm{obs})/\phi_\mathrm{obs}=1/P-1$ (being $\phi_\mathrm{obs}=\phi_\mathrm{true}-(1-P)\phi_\mathrm{true}=P\phi_\mathrm{true}$)
, is therefore smaller than the considered minimum relative error of $\xi=20\%$ on the flux, when only $\gamma$-ray flux data for energies below 100~TeV are used. This applies in the following to the analysis of data for Vela~Jr., RX~J1713.7$-$3946, the GC region, and HESS~J1702$-$420A. As argued in Sec.~\ref{boomerang_sec}, Sec.~\ref{MGRO_section} and Sec.~\ref{sec_unident}, the effect of pair creation can also be neglected for the considered data from the sources LHAASO~J2226$+$6057, MGRO~J1908$+$06 and LHAASO~J2108$+$5157, respectively.

\subsection{Rejecting Pevatron hypotheses: The Supernova Remnants RX~J1713.7$-$3946 and Vela Junior}
\label{pts_ver}

\begin{figure*}
\centering
\includegraphics[width=17cm]{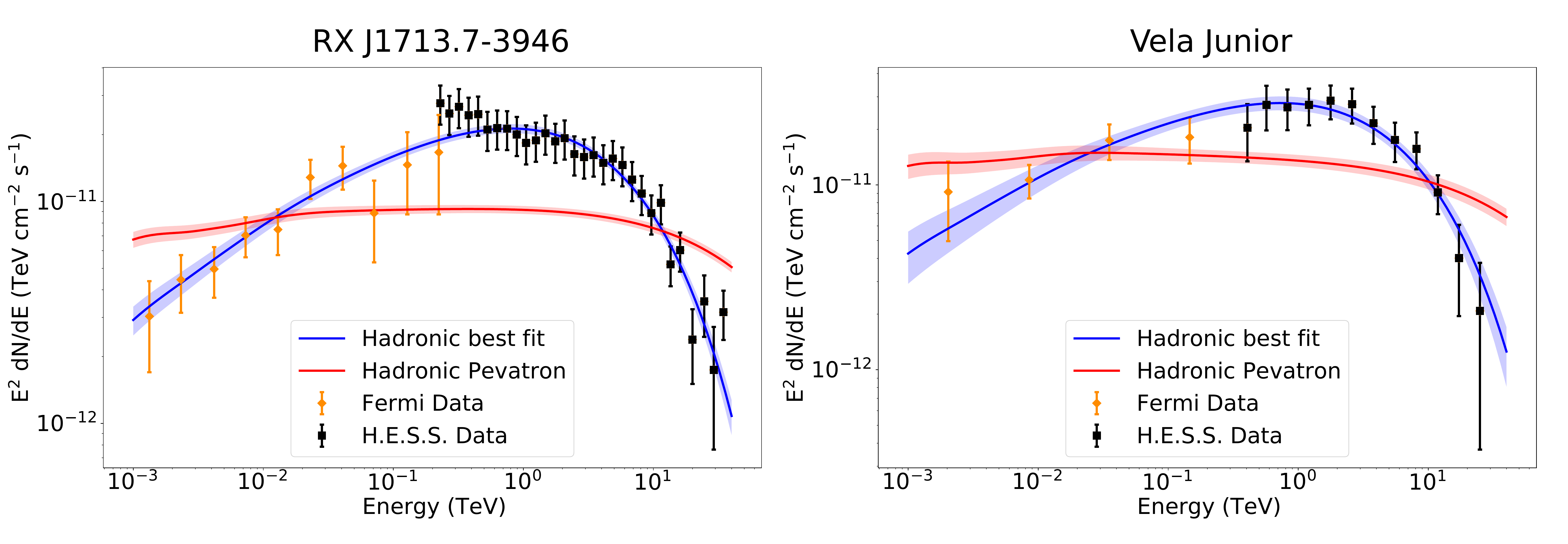}
\caption{$\gamma$-ray spectra of RX~J1713.7$-$3946 (left panel) and Vela~Jr. (right panel). The data are obtained from \protect\cite{hess_1713} for RX~J1713.7$-$3945 and from \protect\cite{vela_hess,fermi_vela} for Vela~Jr. The error bars are derived as detailed in Sec.~\ref{data_analysis} for a minimum relative error of $\xi$=20\%. The best-fit hadronic emission models are shown as blue lines, while best-fit reference Pevatron models, where the cutoff energy of the hadronic particle population is fixed at 1~PeV, are shown in red. Shaded regions indicate 68\% CL uncertainties on the best-fit flux prediction, derived from the propagation of the errors inferred from the respective parameter fit to the shown data, considering the parameter covariance.}
\label{fig_rxj_velajr_joint}
\end{figure*}

\begin{table*}
\caption{RX~J1713.7$-$3946 and Vela~Jr.~analysis results inferred from different energy ranges: H and F in the instrument data column refer to the data derived from observations with H.E.S.S.~(H) and Fermi~(F), respectively. Reference values, on which conclusions are based in this work, are highlighted in bold. All other values are reported to illustrate the dependence of the result on systematic effects or on data selection. $LL_\mathrm{cut,\,p}$ denotes the $95\%$ CL lower limit on the hadronic energy cutoff $E_\mathrm{cut,\,p}$, $\xi$ is the minimal relative flux error as defined in Sec.~\ref{data_analysis}, $S_\mathrm{PTS}$ is the significance of the PTS, $\Gamma_\mathrm{P}$ is the best fit spectral index of the hadron population and the p-value refers to a $\chi^2$ goodness--of--fit test between the data and the best fitting model.}

\label{table_rxj}     
\centering                        
\begin{tabular}{c c c c c c c c c}       
\hline\hline                 
Source & Instrument & Energy & $\xi$ & $\mathrm{S}_\mathrm{PTS}$ & $\Gamma_\mathrm{P}$ & $E_\mathrm{cut,\, p}$ & $LL_\mathrm{cut,p}$ & p-value \\
Name   & Data       & Range (TeV) & ($\%$) & &  & (TeV) & (TeV) & ($\%$) \\ 
\hline       
RX J1713 & H  & [0.2, 34.6] & 0       & -9.9  & 1.99 $\pm$ 0.04 & 105 $\pm$ 16 & 82 & 0   \\
RX J1713 & H  & [0.2, 34.6] & 10 & -7.5  & 1.98 $\pm$ 0.07 & 92 $\pm$ 18  & 67 & 31 \\
RX J1713 & H  & [0.2, 34.6] & 20 & -5.2  & 1.98 $\pm$ 0.11 & 81 $\pm$ 23  & 52 & 97 \\
RX J1713 & H+F & [0.0013, 34.6] & 0       & -28.9 & 1.76 $\pm$ 0.02 & 59 $\pm$ 4   & 52 & 0 \\
RX J1713 & H+F & [0.0013, 34.6] & 10 & -21.5 & 1.72 $\pm$ 0.03 & 53 $\pm$ 5   & 46 & 12 \\
\textbf{RX J1713} & \textbf{H+F} & \textbf{[0.0013, 34.6]} & \textbf{20} & \textbf{-14.5} & \textbf{1.70 $\pm$ 0.04} & \textbf{49 $\pm$ 6}   & \textbf{40} & \textbf{93} \\
\hline                      
Vela Jr. & H & [0.4, 25.0]& 20  & -4.6  & 0.95 $\pm$ 0.63 & 21 $\pm$ 10 & 13 & 100 \\
\textbf{Vela Jr.} & \textbf{H+F} & \textbf{[0.002, 25.0]} & \textbf{20} & \textbf{-7.2}  & \textbf{1.71 $\pm$ 0.08} & \textbf{47 $\pm$ 11} & \textbf{33} & \textbf{97} \\
\hline  
\hline   
\end{tabular}
\end{table*}

RX~J1713.7$-$3946 and Vela Junior are two sources associated with shell-type $\gamma$-ray emitting SNRs.~Despite the constraints on the mean target gas density, purely hadronic emission models as described in Sec.~\ref{data_analysis} are used in the following to model the $\gamma$-ray emission detected from these two SNRs. This is motivated by the putative presence of dense matter clumps in the remnants' surroundings, as detailed in \cite{vela_hess} for Vela~Jr. and in \cite{hadronic_1713, celli_1713} for RX~J1713.7$-$3946.

Figure~\ref{fig_rxj_velajr_joint} shows the $\gamma$-ray spectral data for the two remnants from \cite{hess_1713} for RX~J1713.7$-$3946, and from \cite{vela_hess, fermi_vela} for Vela~Jr. A minimum relative flux error of $\xi=20\%$ is assumed for all flux points seen in Fig.~\ref{fig_rxj_velajr_joint}. The best fit $\gamma$-ray spectra resulting from the assumed hadronic emission model are shown as blue solid lines, while the red lines, shown for comparison, are the best fit $\gamma$--ray spectra when the particle population energy cutoff $E_\mathrm{cut,\,p}$ is fixed to $1$~PeV, i.e. when the sources are modeled as Pevatrons. The figure shows that the fits of the data within a Pevatron model are clearly disfavoured. 

Analysis results obtained for these two remnants are summarized in Tab.~\ref{table_rxj}.~The results for RX~J1713.7$-$3946 are shown in the first six rows of Tab.~\ref{table_rxj}, which differ in the analyzed energy interval and systematic errors taken into account. In the analysis summarized with 20$\%$ systematics (3$^{\text{rd}}$~row), where only data from H.E.S.S. is used, $S_\mathrm{PTS}=-5.2$ is inferred. This result already corresponds to a rejection of the Pevatron hypothesis for RX~J1713.7$-$3946 within the considered hadronic emission model with a significance greater than the $5\sigma$~level. A more robust rejection of the Pevatron hypothesis with a significance of $S_\mathrm{PTS}=-14.5$ is possible when data from Fermi is considered in addition to data from H.E.S.S. (6$^{\text{th}}$ row). As it can be seen from the table, the level of systematics has a strong influence on the obtained $S_\mathrm{PTS}$ values, reflecting in general their importance for the search of Galactic Pevatrons.  

A preference for a break in the energy spectrum of the hadronic particle population for RX~J1713.7$-$3946 is found in \cite{hess_1713}. Arguments for the presence of a hadronic energy break as a result of dense clumps in the remnants environment are discussed in \cite{hadronic_1713}, following \cite{zirkashvili_1713} and \cite{inoue_1713}. Assuming a hadronic particle population with an energy break at $E_\mathrm{break}=1.4$~TeV, the best-fit values for the hadronic energy cutoff $E_\mathrm{cut,\,p}$ and the two spectral indices at energies below and above the energy break found in \cite{hess_1713} are confirmed within errors for $\xi=20\%$ when data from H.E.S.S. and Fermi are fit jointly. Additionally, the Pevatron hypothesis can still be rejected with a significance of $S_\mathrm{PTS}=-6.4$.

Similarly, in the case of Vela~Jr., the addition of data acquired with Fermi allows increasing the significance of the Pevatron hypothesis rejection from $-4.6\sigma$, when only H.E.S.S. data with minimal relative flux error $\xi=20\%$ are considered, to $-7.2\sigma$.~The best-fit values for $\Gamma_\mathrm{p}$ and $E_\mathrm{cut,\,p}$ derived in \cite{vela_hess} for Vela~Jr agree within systematics with the values listed in the last row of Tab. \ref{table_rxj}.

As discussed, due to their age and the presence of a cutoff at TeV energies in the $\gamma$-ray spectrum, Vela~Jr. and RX~J1713.7$-$3946 are typically not believed to be Pevatrons at present times within simple hadronic models. The PTS method confirms this idea with high statistical significance, and, moreover, can quantify the significance of rejection in a straightforward way. A different but very important question is whether these sources were Pevatrons earlier on during their evolution. If this were the case, signatures of the past acceleration of particles to PeV energies might be possible to find by looking at clouds in the source vicinity \citep[e.g.][]{GabiciAharonian07}. Using models for particle acceleration throughout the history of the respective source and particle propagation in the source vicinity, the PTS can also be used to investigate these questions. Appropriate data to carry out such a study will become available with the upcoming generation of high sensitivity, and especially high angular resolution IACTs.

\subsection{The Galactic Center Region}
\label{gc_region}

Observations of the region around the center of the Galaxy across the electromagnetic spectrum have revealed a very complex astrophysical environment.~The compact radio source Sagittarius~A* (Sgr~A*) is found to be spatially coincident with the dynamic center of the Galaxy, and is frequently associated with a supermassive black hole \citep{sgra_bh}. Observations of this region with the MeerKAT radio telescope were discussed in \cite{meerkat} and revealed many SNR structures which can act as potential CR accelerators.~The possible presence of a Galactic Pevatron in this region is discussed in \cite{hess_gc_pevatron}. A review of the research status and further references can be found in \cite{gc_bh_review} and, specifically for the $\gamma$-ray emission from the Galactic Center (GC) region, in \cite{gc_gamma_review}.

\begin{figure*}
\centering
\includegraphics[width=17cm]{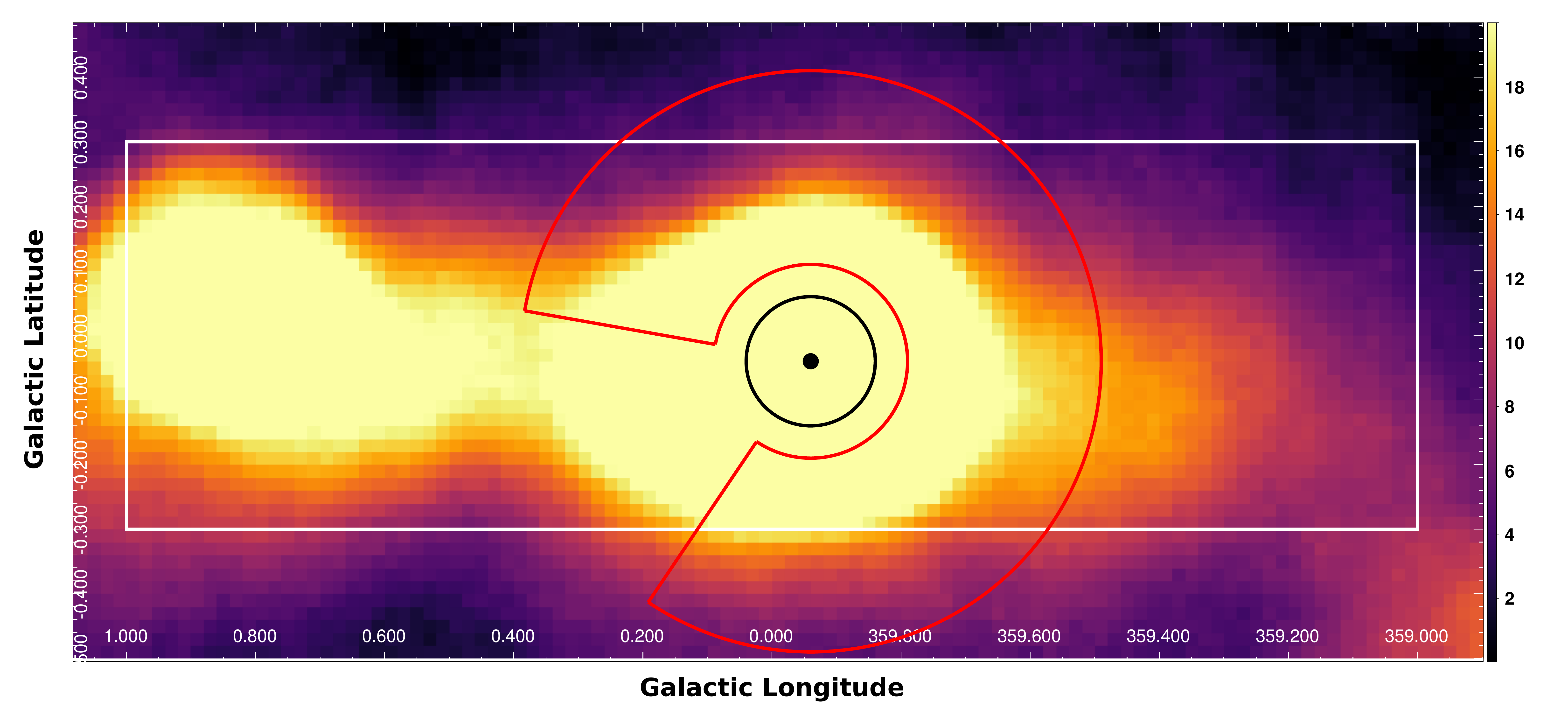}

\caption{Significance map of the GC region extracted from the HGPS catalog \citep{hgps} for the correlation radius of 0.2$^{\circ}$. The black point indicates the position of the compact object Sgr A*. The black circle, the red annulus, and the white rectangle show the data extraction regions used for reconstructing the $\gamma$-ray spectra of HESS~J1745$-$290, the Pacman, and the Galactic ridge regions, respectively. To enhance the visibility of the significant emission originating from the Galactic ridge region, the map is saturated at a level of 20$\sigma$.}
\label{signifmap_gc}
\end{figure*}

The following discussion is limited to VHE $\gamma$-ray data above energies of $\sim$100~GeV, where measurements with multiple instruments and independent data analyses are publicly available.~The analysis of data acquired at energies below $100$~GeV with the Fermi satellite would require a careful consideration of large systematic errors \citep{fermi_sgra} and the putative 'GeV excess' \citep{fermi_gc}, therefore it is not included in the analysis.

\begin{figure*}
\centering
\includegraphics[width=15cm]{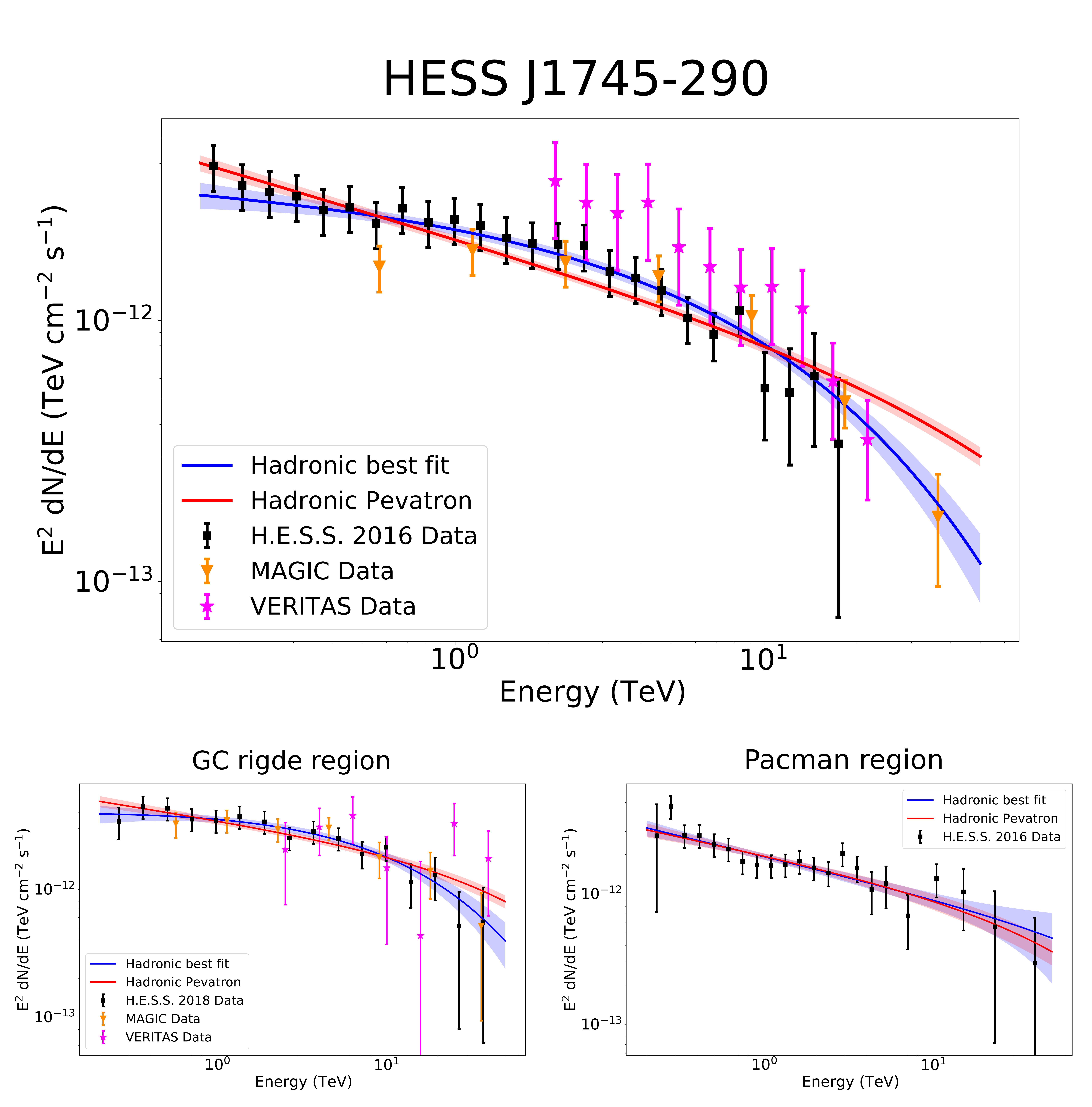}
\caption{Upper panel: Spectral $\gamma$-ray data inferred by \protect\cite{hess_gc_pevatron, veritas_gc, magic_diffuse_gc} from observations with HESS, MAGIC, and VERITAS for the pointlike $\gamma$-ray source HESS~J1745$-$290, which is spatially coincident with the radio source Sgr~A*. Lower panels: Spectral $\gamma$-ray data from \protect\cite{hess_gc_2018, veritas_gc, magic_diffuse_gc} for the diffuse emission from the 'GC ridge' region (left panel) and \protect\cite{hess_gc_pevatron} for the 'Pacman' region (right panel).~The spectrum extraction regions are shown in Fig.~\ref{signifmap_gc}.~The systematic error is treated as described in Sec.~\ref{data_analysis}. For HESS and MAGIC data, $\xi=20\%$ is assumed while spectral data derived from VERITAS observations are analyzed with $\xi=40\%$ (see discussion in the text).~The best-fit hadronic emission models are shown as blue lines, while the best-fit reference Pevatron models, where the cutoff energy of the hadronic particle population is fixed at $1$~PeV, are shown as solid red lines. Shaded regions indicate $68\%$~CL uncertainties.} 
\label{fig_gc}
\end{figure*}

Spectral data from three different regions, as shown in Fig.~\ref{signifmap_gc}, are in the following considered.~The first region is the pointlike source HESS~J1745$-$290, shown with the black circle in Fig.~\ref{signifmap_gc}, and frequently associated with Sgr~A*, although other counterparts are also being discussed \citep{hess_gc_pevatron}. Spectral data for this source are available from three instruments \citep{hess_gc_pevatron, veritas_gc, magic_diffuse_gc} and shown in the upper panel of Fig. \ref{fig_gc}.

\begin{table*}
\caption{Analysis results were obtained for the three different Galactic Center sub-regions shown in Fig.~\ref{signifmap_gc}. Columns are explained in the caption of Tab.~\ref{table_rxj}, except for the letter 'M' in the instrument data column which refers to data from observations with MAGIC. The best fit hadronic cutoff energy, $E_\mathrm{cut,\,p}$, is only given when its relative error is less than $100\%$.}
\label{table_gc}     

\begin{tabular}{c c c c c c c c c}       
\hline\hline\hline
Source & Instrument & Energy & $\xi$ & $\mathrm{S}_\mathrm{PTS}$ & $\Gamma_\mathrm{P}$ & $E_\mathrm{cut,\,p}$ & $LL_\mathrm{cut,\,p}$ & p-value \\
Name   & Data       & Range (TeV) & ($\%$) &  &  & (TeV) & (TeV) &  ($\%$)\\ 
\hline\hline\hline                
HESS~J1745$-$290 & H   & [0.16, 17.4] & 0.       & -4.4  & 2.14 $\pm$ 0.05 & 99 $\pm$ 31 & 62 & 31   \\
HESS~J1745$-$290 & H   & [0.16, 17.4] & 20 & -2.7  & 2.12 $\pm$ 0.13 & 78 $\pm$ 41 & 37 & 100   \\
\hline                        
HESS~J1745$-$290 & M   & [0.57, 36.5] & 0       & -5.3  & 1.51 $\pm$ 0.26 & 50.0 $\pm$ 18 & 27 & 97   \\
HESS~J1745$-$290 & M   & [0.57, 36.5] & 20 & -3.6  & 1.49 $\pm$ 0.40 & 48 $\pm$ 24 & 22 & 100   \\
\hline                                               
HESS~J1745$-$290 & H+M & [0.16, 36.5] & 0       & -6.0  & 2.34 $\pm$ 0.01 & 135 $\pm$ 31 & 94 & 0   \\
HESS~J1745$-$290 & H+M & [0.16, 36.5] & 20 & -3.7  & 2.41 $\pm$ 0.03 & 119 $\pm$ 46 & 67 & 96   \\
\hline                                             
HESS~J1745$-$290 & H+M+V & [0.16, 36.5] & 0       & -6.7  & 2.16 $\pm$ 0.03 & 147 $\pm$ 29 & 108 & 0   \\
\textbf{HESS~J1745$-$290} & \textbf{H+M+V} & \textbf{[0.16, 36.5]} & \textbf{H+M: 20} & \textbf{-4.1}  & \textbf{2.13 $\pm$ 0.09} & \textbf{112 $\pm$ 38}  & \textbf{67} & \textbf{94}   \\
 &  & & \textbf{V: 40} &   &  &   & &    \\
\hline\hline\hline                    
GC Pacman & H & [0.23, 39.6] & 0       & 0.5 & 2.37 $\pm$ 0.09 & Signif. $\ll 1\sigma$ & 185 & 68   \\
\textbf{GC Pacman} & \textbf{H} & \textbf{[0.23, 39.6]} & \textbf{20 } & \textbf{0.4} & \textbf{2.38 $\pm$ 0.09} & \textbf{Signif. $\ll 1\sigma$} & \textbf{172} & \textbf{82}   \\
\hline\hline\hline                       
GC Ridge & H & [0.26, 37.1] & 0       &  -2.1 & 2.08 $\pm$ 0.13 & 160 $\pm$ 98 & 66 & 94 \\
GC Ridge & H & [0.26, 37.1] & 20 &  -2.0 & 2.03 $\pm$ 0.16 & 137 $\pm$ 90 & 53 & 99 \\
\hline      
GC Ridge & M & [0.56, 36.1] & 0       &  -1.7 & 1.81 $\pm$ 0.39 & 97 $\pm$ 90 & 23 & 94  \\
GC Ridge & M & [0.56, 36.1] & 20 &  -1.7 & 1.81 $\pm$ 0.39 & 98 $\pm$ 91 & 23 & 95 \\
\hline    
GC Ridge & H+M & [0.26, 37.1] & 0       & -2.5 & 2.06 $\pm$ 0.11 & 158 $\pm$ 82 & 74 & 99  \\
GC Ridge & H$+$M & [0.26, 37.1] & 20 & -2.5 & 2.01 $\pm$ 0.15 & 135 $\pm$ 73 & 60 & 100  \\
\hline
GC Ridge & H+M+V & [0.26, 39.8] & 0       & -2.4 & 2.08 $\pm$ 0.11 & 179 $\pm$ 93 & 83 & 96 \\
\textbf{GC Ridge} & \textbf{H+M+V} & \textbf{[0.26, 39.8]} & \textbf{H+M: 20} & \textbf{-2.3} & \textbf{2.03 $\pm$ 0.14} & \textbf{157 $\pm$ 87} & \textbf{69} & \textbf{100} \\
 & & & \textbf{V: 40} &  &  &  & &  \\
\hline\hline\hline                         
\end{tabular}
\end{table*}

In addition to the pointlike source HESS~J1745$-$290, the significant detection of diffuse $\gamma$-ray emission around the GC is reported in \cite{hess_diffuse_det}.~Two different sub-regions for the diffuse $\gamma$-ray emission in the vicinity of the GC are considered in the following.~The first sub-region is the 'GC ridge', defined by longitude $|l|<1^\circ$ and latitude $|b|<0.3^\circ$, excluding known $\gamma$-ray sources. This region is shown by the white rectangle in Fig.~\ref{signifmap_gc}.~Spectral data for the 'GC ridge' region are reported in \cite{hess_gc_2018, veritas_gc, magic_diffuse_gc}, and shown in the lower left panel of Fig.~\ref{fig_gc}.~The second sub-region is the 'GC~Pacman', defined in \cite{hess_gc_pevatron} as the annulus around the GC with inner and outer radii of $0.15^\circ$ and $0.45^\circ$ respectively, excluding again known $\gamma$-ray sources. 
Spectral $\gamma$-ray data for this sub-region, shown by the red annulus in Fig.~\ref{signifmap_gc}, are discussed in \cite{hess_gc_pevatron}, and shown in the lower right panel of Fig.~\ref{fig_gc}.~Based on an inferred lower limit of $\sim$400~TeV at $95\%$~CL\footnote{In \cite{porterPacman}, the $95\%$~CL lower limit of 1 PeV is derived for this region taking into account Galactic absorption effects.}, the possible presence of a Pevatron in this sub-region is discussed in \cite{hess_gc_pevatron}.

Empirically, the diffuse $\gamma$-ray emission in the vicinity of the GC exhibits a strong spatial correlation with molecular clouds \citep{hess_diffuse_det}, which suggests a hadronic origin.~A connection between the diffuse $\gamma$--ray emission observed towards the vicinity of the GC and previous phases of enhanced acceleration of hadrons by the SMBH associated with Sgr~A* is, for example, discussed in \cite{hess_gc_pevatron}. An alternative model, where young stellar clusters in the vicinity of the GC accelerate hadrons, is presented in \cite{pevatron_stars}.~In the following, only pure hadronic models for the diffuse $\gamma$-ray emission from the 'GC ridge' and the 'GC~Pacman' regions as well as the central source HESS~J1745$-$290 are considered. Alternative models for the origin of the diffuse $\gamma$-ray emission and the central source HESS\,J1745$-$290 are summarized in \cite{gc_gamma_review}. Systematic errors on the flux normalization and the spectral index are estimated as $15\%$ and $0.1$, respectively, for spectral data derived from H.E.S.S. observations \citep{hess_gc_pevatron}.~For spectral data derived from observations with VERITAS, a $40\%$ systematic error on both the flux normalization and the spectral index are estimated in \cite{veritas_gc}. In the present analysis, we make the following conservative assumptions: an estimated relative uncertainty $\xi=20\%$ is associated to each data point from H.E.S.S. and MAGIC, while $\xi=40\%$ is assumed for VERITAS data.\\ 
The upper panel of Fig.~\ref{fig_gc} shows the three spectral measurements for the point-like source HESS~J1745$-$290. The spectral data inferred from all different observatories are compatible within the assumed errors.~The fit results of the spectral data to the hadronic emission model described in Sec.~\ref{data_analysis} are summarized in Tab.~\ref{table_gc}. 
HESS and MAGIC data immediately provide a strong indication towards the rejection of the Pevatron hypothesis, both considered separately and in a combined manner. The combination of data from HESS, MAGIC, and VERITAS leads to improved significance of $S_\mathrm{PTS}=-4.1$ within systematic errors, and therefore to a rejection of the Pevatron hypothesis for the central source HESS~J1745$-$290. A significant spectral cutoff feature was detected in the $\gamma$-ray spectrum of HESS J1745$-$290 at about 10 TeV \citep{hess_gc_pevatron}, consequently the $\gamma$-ray emission is not expected to be the result of a Pevatron activity. ~The PTS analysis of the region can confirm this result, providing a quantitative rejection level of the Pevatron hypothesis. 

Table \ref{table_gc} summarizes the results of best-fit hadronic $\gamma$-ray emission models to the GC Pacman and the GC Ridge data available.~Again, a minimal relative flux error of $\xi=20\%$ is assumed for data from H.E.S.S. and MAGIC, while $\xi=40\%$ is used for the analysis of spectral data from VERITAS. The PTS leads to $S_\mathrm{PTS}=0.4$ for the GC Pacman region, and to $S_\mathrm{PTS}=-2.3$ for the GC ridge region. Our conclusion is that the data are insufficient to assess the Pevatron hypothesis for both diffuse emission regions based on the PTS. Deeper observations of this region with future instruments, especially at $>$100 TeV energies (i.e.~with the future SWGO experiment), are needed in order to reject or confirm the Pevatron hypothesis for the diffuse $\gamma$-ray emission in the vicinity of the GC.

\subsection{LHAASO J2226$+$6057 and MAGIC Tail Emission: The Boomerang~PWN and SNR~G106.3$+$2.7}
\label{boomerang_sec}

The LHAASO collaboration reported the significant detection of UHE~$\gamma$~rays from the direction of the source LHAASO~J2226$+$6057 at energies above $100$~TeV in~\cite{lhaaso2021}.~Together with previous measurements with different instruments \citep{Ke_FermiNew, veritas2009, tibet_boomerang, milagro2007, milagro2009, magic_tail}, spectral $\gamma$-ray data from GeV to several hundred TeV energies are available for this region. The region was first studied by VERITAS \citep{veritas2009} and secondly by HAWC \citep{hawc_boomerang}. The joint VERITAS-HAWC spectrum can be described well by a power-law with a spectral index of $\sim$2.3, without showing any sign of a spectral cutoff up to 180 TeV. The 90$\%$ C.L. spectral cutoff lower limits on the $\gamma$-ray and proton spectra are found to be 120 TeV and 800 TeV, respectively. Thanks to their improved angular resolution, the recent results from the MAGIC Collaboration \citep{magic_tail} provided for the first time clear evidence for the existence of two emission components in the region, while the data from other experiments did not show any hint for separate components.~The soft component, called 'head', has a spectral index of $\Gamma_{\text{H}}$~=~2.12~$\pm$~0.12, while the spectral index of the hard component, called 'tail', is found to be $\Gamma_{\text{T}}$~=~1.83~$\pm$~0.10. The best-fit positions of the head and tail components can be statistically separated from each other, having their emissions centered at RADEC coordinates of (337$^{\circ}$.13, 61$^{\circ}$.10) and (336$^{\circ}$.72, 60$^{\circ}$.84), respectively, and a spatial extensions of 0.16$^{\circ}$ \citep{magic_tail}. 

\begin{figure*}
\centering
\includegraphics[width=18cm]{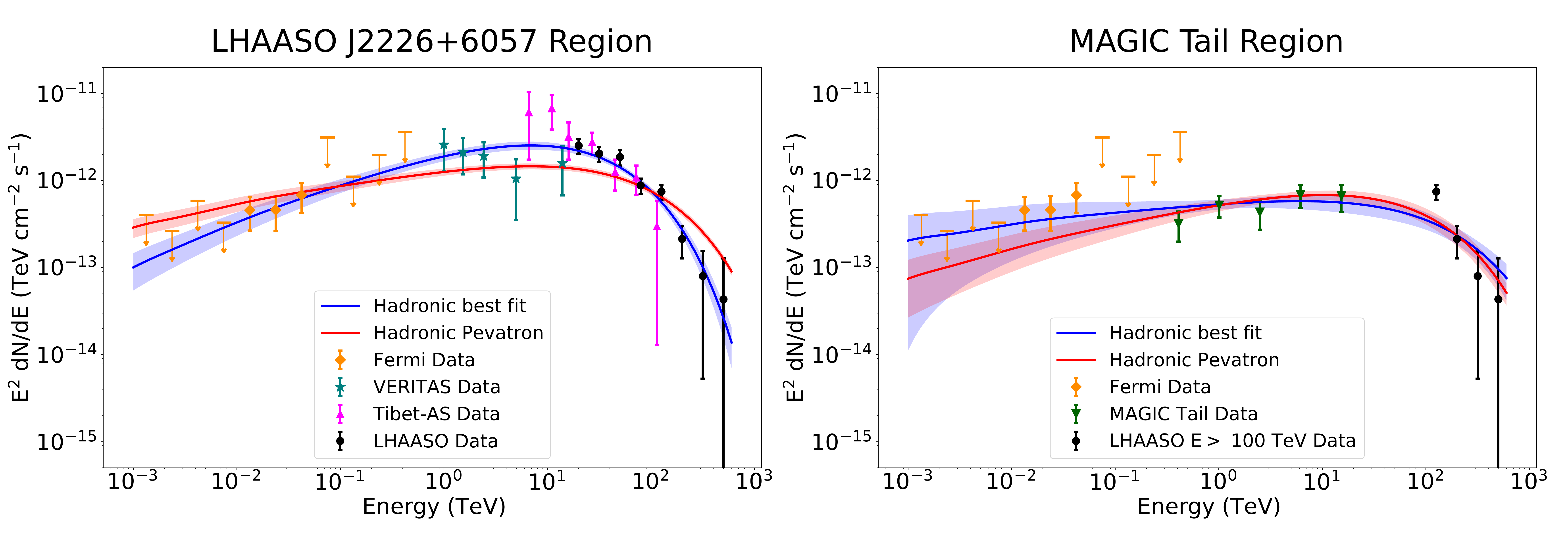}
\caption{Left: Spectral $\gamma$-ray data reported in \protect\cite{Ke_FermiNew} (Fermi), \protect\cite{veritas2009} (VERITAS), \protect\cite{tibet_boomerang} (Tibet~AS$\gamma$) and \protect\cite{lhaaso2021} (LHAASO) for the $\gamma$-ray emission region which contains the Boomerang PWN and SNR~G106.3+2.7. VERITAS data are scaled by a factor of $1.62$ to match the signal region of the Tibet~AS$\gamma$ analysis, as detailed in \protect\cite{tibet_boomerang}. The minimum relative flux error is assumed to be $\xi=20\%$ for all flux points and treated as described in Sec. \protect\ref{data_analysis}. Right: Spectral $\gamma$-ray data reported in \protect\citep{magic_tail} (MAGIC) and \protect\cite{lhaaso2021} (LHAASO). For the LHAASO data, only the spectral flux points above 100 TeV are taken into account. The spectral data shown in Fig.~\ref{fig_boomerang} for Fermi are derived under the assumption of a pointlike source model because an extended source model is not statistically preferred \protect\citep{Ke_FermiNew}. The best-fit hadronic emission models to all data excluding the shown flux limits are shown as blue lines, while the best-fit reference Pevatron models, where the cutoff energy of the hadronic particle population is fixed at $1$~PeV, are shown with the solid red line. Shaded regions indicate $68\%$~CL errors. }
\label{fig_boomerang}
\end{figure*}

\begin{table*}
\caption{Analysis results for the $\gamma$-ray emission region containing the Boomerang PWN and SNR~G106.3$+$2.7. See Tab. \ref{table_rxj} for an explanation of the columns. The letters in the instrument data column indicate different observatories: F (Fermi, \citep{fermi_j2227}), F$_{\text{P}}$ (Fermi pointlike source, \citep{Ke_FermiNew}), F$_{\text{G}}$ (Fermi Gaussian source extension, \citep{Ke_FermiNew}) V (VERITAS), M$_{\text{T}}$ (MAGIC Tail), T (Tibet~AS$\gamma$), L (LHAASO), L$_{\text{100}}$ (LHAASO E$>$ 100 TeV). The best fit hadronic cutoff energy, $E_\mathrm{cut,\, p}$, is only given when its relative error is less than $100\%$.}
\label{table_boomerang}     
\centering           
\begin{tabular}{c c c c c c c c c}       
\hline\hline                 
Source & Instrument & Energy & $\xi$ & $\mathrm{S}_\mathrm{PTS}$ & $\Gamma_\mathrm{p}$ & $E_\mathrm{cut,p}$ & $LL_\mathrm{cut,p}$ & p-value\\
Region & Data       & Range (TeV) & ($\%$) &  &  & (TeV) & (TeV) & ($\%$) \\ 
\hline                        
LHAASO~J2226$+$6057 & F & [0.005, 0.301] & 20 & 0.1  & 1.88 $\pm$ 0.18 & Signif. $\ll 1\sigma$ & 0.2 & 78\\
LHAASO~J2226$+$6057 & V & [1.0, 13.9] & 20 & 0.3  & 2.37 $\pm$ 0.34 & Signif. $\ll 1\sigma$ & 6 & 70 \\
LHAASO~J2226$+$6057 & T & [6.6, 114.0] & 20 & -0.5 & 2.36 $\pm$ 1.02 & Signif. $\ll 1\sigma$ & 46 & 83\\
LHAASO~J2226$+$6057 & L & [20.0, 501.0] & 0               & -2.0 & 1.38 $\pm$ 0.76 & 241 $\pm$ 131 & 124 & 28 \\
LHAASO~J2226$+$6057 & L & [20.0, 501.0] & 20         & -1.6 & 1.46 $\pm$ 0.93 & 256 $\pm$ 174 & 121 & 69 \\
\hline                                            
LHAASO~J2226$+$6057 & F+V+T & [0.005, 114.0] & 20 & -1.5 & 1.77 $\pm$ 0.08 & 437 $\pm$ 195 & 224 & 66 \\
\hline                        
LHAASO~J2226$+$6057 & F+V+L & [0.005, 501.0] & 0           & -5.1 & 1.69 $\pm$ 0.07 & 378 $\pm$ 65 & 285 & 34 \\
LHAASO~J2226$+$6057 & F+V+L & [0.005, 501.0] & 20     & -4.1 & 1.72 $\pm$ 0.07 & 399 $\pm$ 77 & 292 & 71 \\
\hline                        
LHAASO~J2226$+$6057 & F+V+T+L & [0.005, 501.0] & 0    & -5.5 & 1.68 $\pm$ 0.07 & 355 $\pm$ 60 & 268 & 41\\
LHAASO~J2226$+$6057 & F+V+T+L & [0.005, 501.0] & 20   & -4.7 & 1.70 $\pm$ 0.07 & 367 $\pm$ 68 & 271 & 71 \\
LHAASO~J2226$+$6057 & F$_{\text{G}}$+V+T+L & [0.013, 501.0] & 20    & -4.7 & 1.69 $\pm$ 0.08 & 357 $\pm$ 68 & 260 & 63\\
\textbf{LHAASO~J2226$+$6057} & \textbf{F$_{\text{P}}$+V+T+L} & \textbf{[0.013, 501.0]} & \textbf{20} & \textbf{-5.2} & \textbf{1.62 $\pm$ 0.08} & \textbf{327 $\pm$ 60} & \textbf{241} & \textbf{71}\\
\hline                        
\hline                                    
MAGIC Tail & M$_{\text{T}}$ & [0.4, 15.3] & 20  & 0.10 & 1.76 $\pm$ 0.47 & Signif. $\ll 1\sigma$ & 32 & 64 \\
\hline
MAGIC Tail & M$_{\text{T}}$ + L$_{100}$ & [0.4, 501.0] & 20  & -1.0 & 1.51 $\pm$ 0.25 & 619 $\pm$ 281 & 283 & 25 \\
\hline
MAGIC Tail & F$_{\text{P}}$ + M$_{\text{T}}$ & [0.013, 15.3] & 20  & 1.3 & 2.02 $\pm$ 0.07 & Signif. $\ll 1\sigma$ & 466 & 57 \\
MAGIC Tail & F$_{\text{G}}$ + M$_{\text{T}}$ & [0.013, 15.3] & 20  & 1.3 & 2.07 $\pm$ 0.09 & Signif. $\ll 1\sigma$ & 527 & 24 \\
\hline
MAGIC Tail & F$_{\text{G}}$ + M$_{\text{T}}$ + L$_{100}$& [0.013, 501.0] & 20  & 1.0 & 1.95 $\pm$ 0.14 & 1799 $\pm$ 1152 & 669 & 3 \\
\textbf{MAGIC Tail} & \textbf{F$_{\text{P}}$ + M$_{\text{T}}$ + L$_{100}$}& \textbf{[0.013, 501.0]} & \textbf{20}  & \textbf{1.2} & \textbf{1.94 $\pm$ 0.10} & \textbf{1750 $\pm$ 878} & \textbf{817} & \textbf{8} \\
\hline                        
\hline 
\end{tabular}
\end{table*}
Two different astrophysical objects, SNR~G106.3$+$2.7 and the Boomerang~PWN, have been discussed as plausible sources of the observed $\gamma$-ray emission.~The distance to SNR~G106.3$+$2.7 is estimated to be less than 1~kpc~\citep{boomerang_distance}.~As discussed in~\cite{veritas2009}, the VHE emission seen by VERITAS is centered near the peak of a dense~$^{12}$CO~region which suggests a hadronic origin of the emission.~The acceleration of particles by SNR~G106.3$+$2.7 is discussed in \cite{hawc_boomerang}.~However, as for example noted in \cite{alison}, SNR~G106.3$+$2.7 is older than $3.9$~kyrs and therefore unlikely to accelerate particles to PeV energies. An alternative hadronic origin of the emission powered by the Boomerang~PWN is discussed in \cite{fermi_j2227}. The multi-wavelength investigation of the emission from the tail region suggests a hadronic origin, while the nature of the emission mechanism from the head region can be both leptonic or hadronic \citep{magic_tail}. Given the spatial proximity of SNR~G106.3$+$2.7 and following \cite{absorption_vernetto}, the attenuation of the $\gamma$-ray spectrum due to pair creation is expected to be much smaller than $10\%$. Within the assumed systematic error, the effect of $\gamma$-ray attenuation can therefore be neglected. In order to demonstrate the power and effect of resolving source confusion in Pevatron searches, analyses of two different datasets, one for the entire region (LHAASO~J2226$+$6057) covering both the SNR~G106.3$+$2.7 and the Boomerang PWN, and the other for the tail region only, are performed. 

Figure \ref{fig_boomerang} (left) shows $\gamma$-ray data from the entire region including both head and tail regions, together with the best fit hadronic emission model shown in blue.~The data acquired with Fermi \citep{Ke_FermiNew}, VERITAS \citep{veritas2009}, Tibet-AS$\gamma$ \citep{tibet_boomerang} and LHAASO \citep{lhaaso2021} were used for the analysis of this emission region. As discussed in \cite{tibet_boomerang}, spectral data from VERITAS observations in Fig.~\ref{fig_boomerang} are scaled by a factor of $1.62$ to adjust for the differences in the integration radius between the different analyses.~On the other hand, Fig.~\ref{fig_boomerang} (right) shows $\gamma$-ray emission only from the tail region. Energy-dependent morphology investigation of Fermi data shows that the high energy $\gamma$-ray emission above 10 GeV is centered at RADEC coordinates of (336$^{\circ}$.71, 60$^{\circ}$.90) \citep{Ke_FermiNew}, while the UHE emission from the direction of LHAASO J2226+6057 above 100 TeV is centered at RADEC coordinates of (336$^{\circ}$.75, 60$^{\circ}$.95). Fermi and LHAASO emission are therefore found to be coincident with the reported emission from the tail region. Furthermore,  \cite{magic_tail} discussed that the contribution of head emission to the total flux above 10 TeV is below 37.1$\%$.~Using the power-law spectral models for head and tail regions given in \cite{magic_tail}, this contribution can be calculated as 22.6$\%$ above 50 TeV and 19.2$\%$ above 100 TeV. In order to ensure that possible contamination coming from the head region is still within our minimum relative error of $\xi=20\%$, only the LHAASO spectral points above 100 TeV, together with Fermi and MAGIC tail data, are taken into account in the joint fit shown in Fig. \ref{fig_boomerang} (right).

Quantitative results for the fit of the hadronic emission model described in Sec.~\ref{data_analysis} to the available spectral data for the entire region and tail region are summarized in Tab.~\ref{table_boomerang}. For the entire region, assuming a single emission component, the combination of data from Fermi, VERITAS, Tibet-AS$\gamma$, and LHAASO results in $S_\mathrm{PTS}=-5.2$. In this case, it is therefore excluded with a statistical significance of more than $5\sigma$ that the source associated with LHAASO~J2226+6057 is a Pevatron.~The best-fit energy cutoff of the hadronic particle population is $E_\mathrm{cut,\,p}=(327\pm60)$~TeV together with the 95$\%$ CL lower limit of 241 TeV. Table~\ref{table_boomerang} for the LHAASO~J2226+6057 region also highlights the importance of the combination of data over a wide range of energies. With only data from one of the considered experiments, a decision on the Pevatron hypothesis based on the PTS is impossible, while combining the different data sets can results in significant rejection. On the other hand, the fit of the hadronic emission to the available spectral data for the tail region shown in Fig.~\ref{fig_boomerang} (right) results in $S_\mathrm{PTS}=1.2$ and the best-fit energy cutoff of the hadronic particle population is $E_\mathrm{cut,\,p}=(1750\pm878)$~TeV with the 95$\%$ CL lower limit on the hadronic cutoff energy of $\sim$820~TeV, which provides more promising Pevatron picture with respect to joint HAWC and VERITAS analysis. Based on the results obtained from the joint analysis of currently available $\gamma$-ray data for the tail region, it is therefore impossible to decide whether the source is a Pevatron contributing to the CR spectrum above 1 PeV, and further observations are needed. When an extended source model for the data acquired with Fermi is assumed, instead of a pointlike source model, and a hadronic emission model is fitted to otherwise unchanged data, the results obtained both for the entire and tail only regions do not change significantly (see Tab.~\ref{table_boomerang}). 

The importance of improved angular resolution in the hunt for Galactic Pevatrons is demonstrated in light of recent MAGIC results. In the case when source confusion can not be resolved and the emission from the region is assumed to result from a single component (i.e.~LHAASO~J2226+6057), the joint data analysis results in a significant rejection of the Pevatron hypothesis with $S_\mathrm{PTS}=-5.2$. On the contrary, when the source confusion can be resolved with high angular resolution observations and the emission can be separated into two components, the joint analysis leads to $S_\mathrm{PTS}=1.2$ and a lower limit on the cutoff energy is 817~TeV, therefore indicating the source as one of the most intriguing Pevatron candidates.~The future CTA observations of the tail region can indeed provide unprecedented angular resolution together with spectral data, especially between 10~TeV and 100~TeV, and therefore can lead to robust identification of the Pevatron nature of the tail region.

\subsection{The unidentified UHE source: MGRO J1908+06}
\label{MGRO_section}

One of the most promising Pevatron candidates is the unidentified source MGRO J1908+06.~Both the LHAASO and HAWC Collaborations reported significant $\gamma$-ray emission above 100 TeV coming from the direction of this source \citep{lhaaso2021, hawc_j1908}.~Several astrophysical objects in the region can be responsible for the observed $\gamma$-ray emission.~Two pulsars, PSR~J1907+0602 and PSR~J1906+0722, with $\dot{\text{E}}$ values of 2.8$\times$10$^{36}$~erg/s and 1.0$\times$10$^{36}$~erg/s, respectively, can produce leptonic emission. Moreover, there are also two SNRs, SNR G40.5-0.5 and SNR 3C397, and dense molecular clouds located in the emission region.~Especially, the interaction between SNR G40.5-0.5 and dense molecular clouds located around the SNR, with gas densities ranging between [110,~280]~cm$^{-3}$ (for a near kinematic distance of 3.7~kpc) and [260,~660]~cm$^{-3}$ (for a far kinematic distance of 8.7~kpc), can give rise to hadronic emission.~It was discussed in \cite{hawc_j1908} that the multi-wavelength modelling of the emission suggests preferably a leptonic origin, while a hadronic origin cannot be excluded.

\begin{figure}
\centering
\includegraphics[width=10cm]{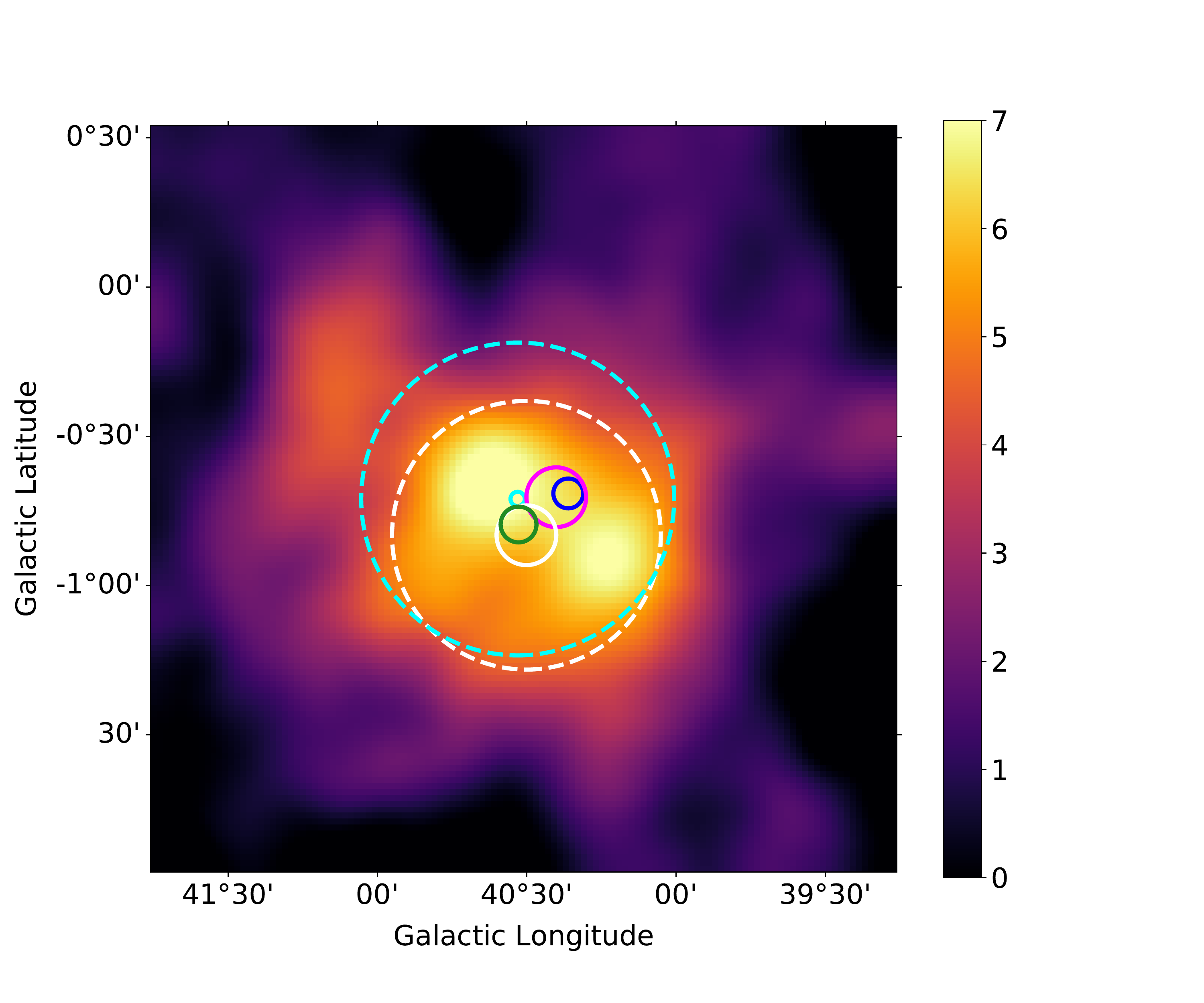}
\caption{Significance map of the MGRO J1908+06 region from the HGPS catalog \citep{hgps} for the correlation radius of 0.2$^{\circ}$.~The 1$\sigma$ statistical error on the best-fit position obtained from Fermi \citep{fermi_J1907}, HESS \citep{hess_J1907}, HAWC (Gauss) \citep{hawc_E56TeV}, HAWC (Diffuse) \citep{hawc_j1908} and LHAASO \citep{lhaaso2021} observations are shown with solid blue, cyan, magenta, green and white circles, respectively. The source extension (given in Gaussian sigma) from HESS and LHAASO observations are shown with dashed cyan (0.524$^{\circ}$) and white (0.45$^{\circ}$) circles, respectively.}
\label{signifmap_j1907}
\end{figure}

\begin{figure*}
\centering
\includegraphics[width=18cm]{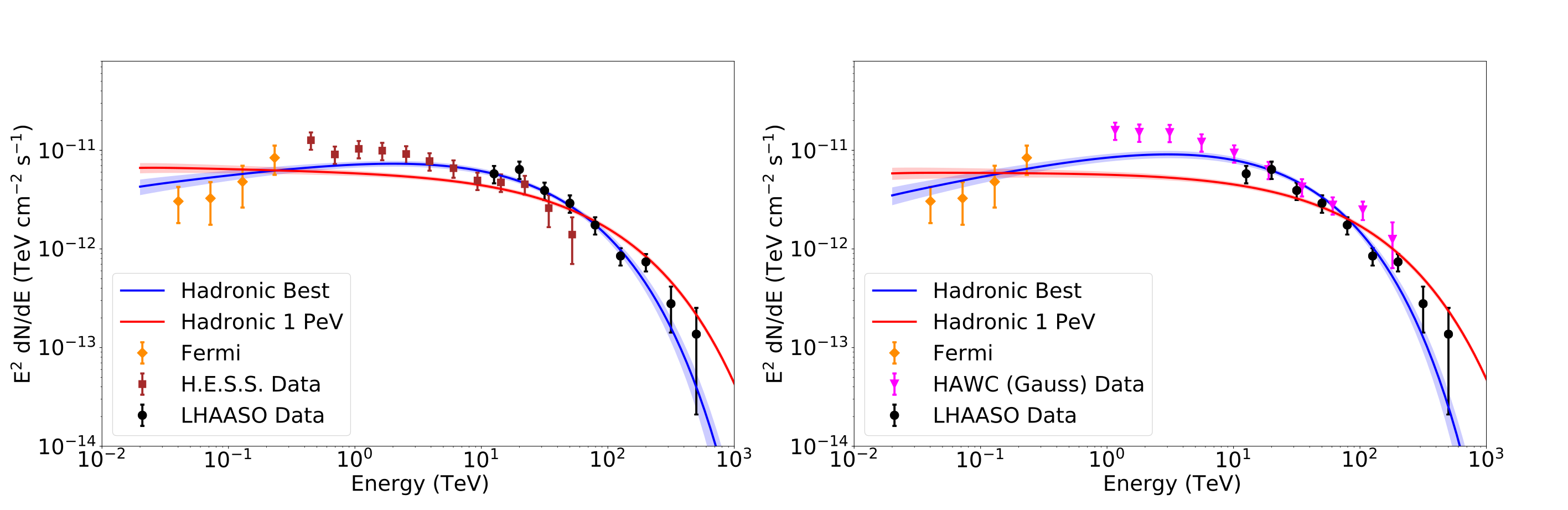}
\includegraphics[width=18cm]{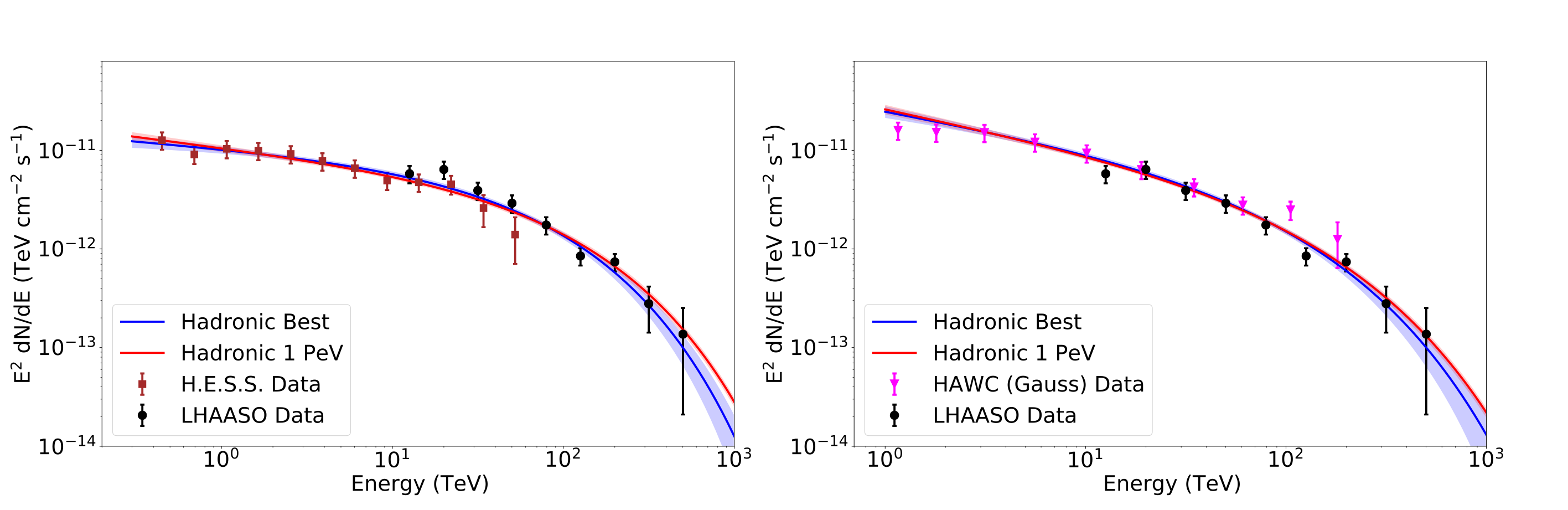}
\caption{Joint $\gamma$-ray spectral data for the MGRO J1908+06 region acquired from Fermi \citep{fermi_J1907}, HESS \citep{hess_J1907}, HAWC (Gauss) \citep{hawc_E56TeV}, HAWC (Diffuse) \citep{hawc_j1908} and LHAASO \citep{lhaaso2021} observations.~Upper panel: Joint spectral data analyses using Fermi and LHAASO data together with H.E.S.S. flux points (left plot, giving $\mathrm{S}_\mathrm{PTS}$ = -5.82$\sigma$) and HAWC (Gauss) flux points (right plot, giving $\mathrm{S}_\mathrm{PTS}$ = -7.15$\sigma$) are shown.~Lower panel: Joint spectral data analyses using LHAASO data together with H.E.S.S. flux points (left plot, giving $\mathrm{S}_\mathrm{PTS}$ = -1.40$\sigma$) and HAWC (Gauss) flux points (right plot, giving $\mathrm{S}_\mathrm{PTS}$ = -1.30$\sigma$) are shown.~The best fits to hadronic models are shown with blue lines and butterflies, while the reference 1 PeV hadronic models are shown with red lines and butterflies. The minimum relative flux error is assumed to be $\xi=20\%$ for all flux points.}
\label{fig_j1908}
\end{figure*}

The $\gamma$-ray data available for this region cover a wide energy range from a few tens of GeV to several hundred TeV, acquired from Fermi \citep{fermi_J1907}, HESS \citep{hess_J1907}, HAWC \citep{hawc_j1908, hawc_E56TeV} and LHAASO \citep{lhaaso2021} observations. The source displays a single component with an extended morphology ($>$0.5$^{\circ}$) in the HE$-$VHE domain, and remains extended even in the UHE domain (0.45$^{\circ}$).~The 1$\sigma$ statistical uncertainties on the best-fit positions derived from different observations are shown in Fig.~\ref{signifmap_j1907}.~One can see from the figure that all best-fit positions are compatible within 3$\sigma$ uncertainties.~In contrast to the case of SNR~G106.3+2.7 discussed in Sect.~\ref{boomerang_sec}, the recent observations taken with HESS telescopes, reaching up to a total live time of 80 h and providing relatively good angular resolution compared to the other experiments (see Fig.~\ref{signifmap_j1907}), were not sufficient to resolve more than a single component or any energy-dependent morphology in the region, leaving the hotspot structures seen in the data still in agreement within uncertainties \citep{hess_J1907}.~Consequently, the connection between the observed GeV and $>$100 TeV emission remains unclear. 

\begin{table*}
\caption{Analysis results for the $\gamma$-ray emission coming from the region MGRO J1908+06. Columns are explained in the caption to Tab.~\ref{table_rxj} except  for the letter indicating the instrument from which the data come: F: Fermi, H: H.E.S.S., HWG: HAWC (Gauss), HWD: HAWC (Diffuse), L:LHAASO.}
\label{table_j1908}     
\centering                        
\begin{tabular}{c c c c c c c c}       
\hline\hline                 
Instrument & Energy & $\xi$  & $\mathrm{S}_\mathrm{PTS}$ & $\Gamma_{p}$ & $E_\mathrm{cut,p}$ & $LL_\mathrm{cut,p}$ & p-value  \\
Data       & Range (TeV) & ($\%$) & &  & (TeV) & (TeV) & ($\%$) \\ 
\hline                        

F   & [0.040, 0.231] & 20 & 0.16  & 1.45 $\pm$ 0.34 & Signif. $\ll 1\sigma$ & 1.1   & 54 \\
H   & [0.449, 52.3] & 20 & -1.31 & 2.13 $\pm$ 0.15 & 306 $\pm$ 212 & 116 & 96 \\
HWG & [1.2, 180.8] & 20 & 0.11  & 2.33 $\pm$ 0.13 & 1093 $\pm$ 886 & 376 & 89 \\
HWD & [1.2, 167.0] & 20 & -0.49 & 2.34 $\pm$ 0.15 & 672 $\pm$ 489  & 258 & 54 \\
L   & [12.6, 501.0] & 20 & -0.05 & 2.47 $\pm$ 0.31 & 965 $\pm$ 730  & 318 & 59 \\
\hline                        
F+H   & [0.040, 52.3] & 20 & -5.31 & 1.65 $\pm$ 0.11 & 93  $\pm$ 31 & 55  & 59 \\
F+HWG & [0.040, 180.8] & 20 & -4.89 & 1.64 $\pm$ 0.09 & 179 $\pm$ 49 & 115 & 7 \\
F+HWD & [0.040, 167.0] & 20 & -6.21 & 1.50 $\pm$ 0.11 & 123 $\pm$ 27 & 85  & 2 \\
F+L   & [0.040, 501.0] & 20 & -3.89 & 1.83 $\pm$ 0.07 & 355 $\pm$ 79 & 250 & 54 \\
\hline  
\textbf{H+L} & \textbf{[0.449, 501.0]} & \textbf{20} & \textbf{-1.4} & \textbf{2.20 $\pm$ 0.07} & \textbf{686 $\pm$ 175} & \textbf{458} & \textbf{85} \\
\textbf{HWG+L} & \textbf{[1.2, 501.0]} & \textbf{20} & \textbf{-1.3} & \textbf{2.33 $\pm$ 0.09} & \textbf{671 $\pm$ 196} & \textbf{427} & \textbf{60} \\
HWD+L & [1.2, 501.0] & 20 & -0.8 & 2.44 $\pm$ 0.08 & 768 $\pm$ 239 & 476 & 31 \\
\hline  
\textbf{F+H+L}   & \textbf{[0.040, 501.0]} & \textbf{20} & \textbf{-5.82} & \textbf{1.89 $\pm$ 0.05} & \textbf{344 $\pm$ 57} & \textbf{262} & \textbf{8} \\
\textbf{F+HWG+L} & \textbf{[0.040, 501.0]} & \textbf{20} & \textbf{-7.15} & \textbf{1.78 $\pm$ 0.06} & \textbf{259 $\pm$ 41} & \textbf{200} & \textbf{1} \\
F+HWD+L & [0.040, 501.0] & 20 & -7.27 & 1.76 $\pm$ 0.06 & 239 $\pm$ 39 & 184 & $\ll$ 0 \\
\hline                        
F+H+HWG+L & [0.040, 501.0] & 20 & -7.23 & 1.88 $\pm$ 0.04 & 320 $\pm$ 46 & 252 & 1 \\
F+H+HWG+L & [0.040, 501.0] & 80 & -4.19 & 1.84 $\pm$ 0.07 & 352 $\pm$ 74 & 252 & 99 \\
F+H+HWD+L & [0.040, 501.0] & 20 & -7.25 & 1.87 $\pm$ 0.04 & 316 $\pm$ 45 & 250 & $\ll$ 0 \\

\hline                        
\hline                                    
\end{tabular}
\end{table*}

In this section, two different assumptions are made in order to investigate the Pevatron nature of the observed emission, assuming pure hadronic origin. The first approach assumes that there is only one source in the region, therefore the Fermi GeV and UHE emission have the same origin, while the second approach assumes that there are two different origins responsible for the GeV and UHE emission.~Table~\ref{table_j1908} summarizes the fit results obtained from different combinations of the available spectral data to the hadronic emission model. For the former case, assuming a single origin, joint analyses of combined Fermi, HESS (or HAWC), and LHAASO data result in significant rejection of the Pevatron hypothesis, regardless of whether HESS or HAWC data are used (see Tab.~\ref{table_j1908}). Figure~\ref{fig_j1908} (top) shows available joint spectral $\gamma$-ray data using HESS (left) and HAWC (right) observations, giving $\mathrm{S}_\mathrm{PTS}$ of -5.82$\sigma$ and -7.15$\sigma$, respectively. On the other hand, assuming a common origin for the VHE and UHE emission and a different origin for the GeV emission, joint analysis of combined HESS (or HAWC) and LHAASO data does not allow one to reject or accept the Pevatron hypothesis, resulting in insignificant $\mathrm{S}_\mathrm{PTS}$ of -1.40$\sigma$ and -1.30$\sigma$, respectively, as shown in Fig.~\ref{fig_j1908} (bottom). As it was shown in Extended Data Fig.~6 of \cite{lhaaso2021}, the attenuation of the $\gamma$-ray spectrum of 
LHAASO~J1908$+$0621 due to pair creation is expected to be smaller than 20$\%$ for the energies below $\sim$600 TeV, which is compatible with the assumed systematic errors, and can therefore be neglected.

Joint analyses of the currently available $\gamma$-ray data from this region show no hint of the acceleration of hadrons well beyond 1 PeV energies, consequently no signature for a possible contribution to the 3 PeV knee seen in the CR spectrum could be found in the data. However, given the number of hotspot structures seen in HESS observations of this region, it is possible that there are at least two (or more) sub-components contributing to the observed $\gamma$-ray emission. Similar to the case of SNR~G106.3+2.7 discussed in Sect.~\ref{boomerang_sec}, it is likely that at least one of possible sub-components can have hard spectra reaching up to energies above 100 TeV, producing UHE $\gamma$-ray emission detectable by LHAASO. Deep observations of this region with the future CTA experiment, covering energies from a few tens of GeV up to a few hundred TeV and with its superior angular resolution, can shed light on whether there is more than one source in the region, and pinpoint the origin of the UHE $\gamma$-ray emission.


\subsection{Two unidentified sources: LHAASO~J2108$+$5157 and HESS~J1702$-$420A}
\label{sec_unident}

\begin{figure*}
\centering
\includegraphics[width=18cm]{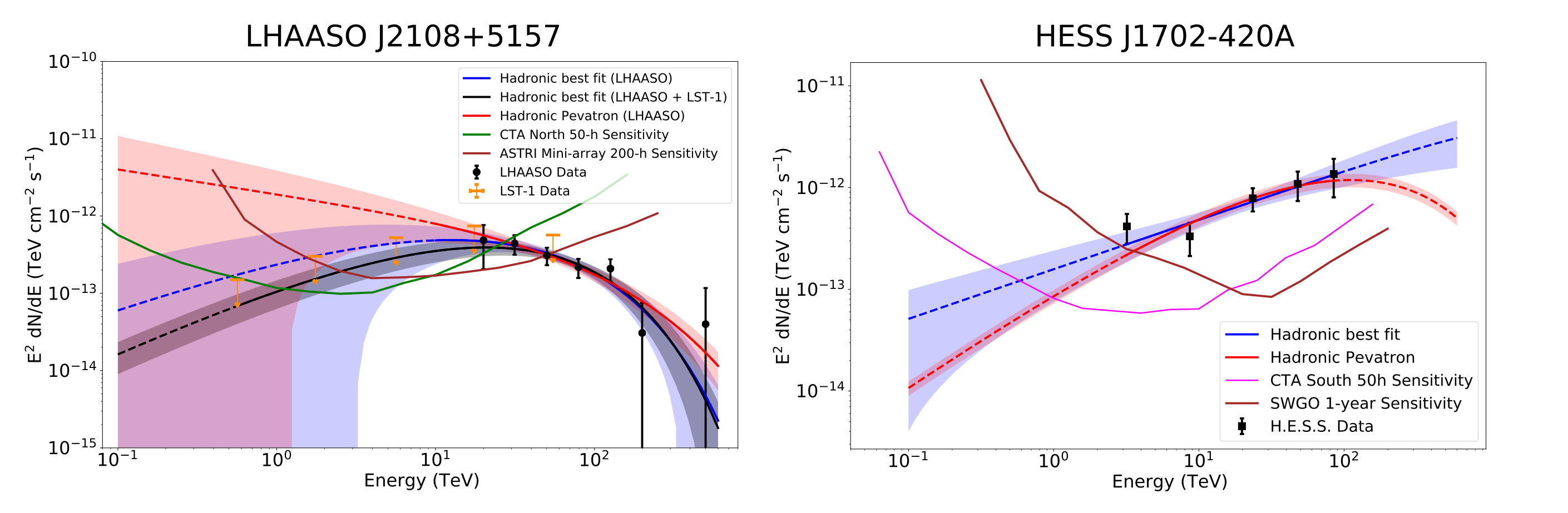}
\caption{Left: Spectral data for LHAASO~J2108+5157 from \protect\cite{lhaaso_new_source} as black points with errors and $95\%$ CL upper limits on the $\gamma$-ray flux from LST-1 \protect\citep{lstJ2108} in orange. Also shown is the sensitivity of the ASTRI Mini-Array \protect\citep{astri_sens} as a brown solid line and planned Northern CTA observatory as a green solid line \protect\citep{prod5}. The black line is a best-fit hadronic emission model to the data from LHAASO which does not violate the shown flux limits from LST-1. Right: Spectral data in the energy range between $3.2$~TeV and $84.8$~TeV for HESS~J1702--420A from \protect\cite{j1702} in black with 1$\sigma$ error bars. The 1-year reference configuration sensitivity of SWGO (brown) is taken from https://github.com/harmscho/SGSOSensitivity and the CTA South 50h sensitivity (magenta) is taken from \protect\cite{prod5}. For all flux data points a minimum relative flux error of $\xi=20\%$ is assumed. The best fit hadronic emission models are shown as solid blue lines. The best-fit reference Pevatron models, where the cutoff energy of the hadronic particle population is fixed at $1$~PeV, are shown in solid red. Dashed lines indicate extrapolations of the corresponding best-fit models, while the shaded regions indicate $68\%$~CL error bands.}        
\label{j2108_single}
\end{figure*}

Recent~analyses of data acquired respectively with the~LHAASO and~HESS observatories resulted in the detection of two previously unknown $\gamma$-ray sources, LHAASO~J2108$+$5157~\citep{lhaaso_new_source} and HESS~J1702$-$420A~\citep{j1702}. The latter source was detected as a sub-component of the bright H.E.S.S. source HESS~J1702$-$420 \citep{j1702_2009}. The $\gamma$-ray energy spectra of both sources are compatible with power-law models, showing no clear indications for spectral $\gamma$-ray cutoff up to at least several tens of TeV. Therefore, both sources are considered as potential Pevatron candidates.

A spatial correlation with molecular clouds, and consequently a hadronic origin of the observed $\gamma$-ray emission, is plausible for LHAASO~J2108$+$5157~\citep{lhaaso_new_source, j2108_new_mc}. Based on work presented by \cite{pevatron_stars}, \cite{lhaaso_new_source} discuss the possibility that the $\gamma$-ray emission from LHAASO~J2108$+$5157 may result from the interactions of hadrons accelerated in young stellar clusters. Figure \ref{j2108_single} (left) shows the available spectral data for LHAASO~J2108$+$5157. The analysis of these data results in an insignificant PTS with $S_\mathrm{PTS}=-0.6$, and a $95\%$~CL lower limit of $102$~TeV on $E_\mathrm{cut,p}$, when a minimum relative flux error of $\xi=20\%$ is considered. Based on the currently available LHAASO data only, it is impossible to decide whether the source is a Pevatron or not, and further observations, especially at energies lower than 10~TeV, are needed. 

As a result of observations with the single Large~Size~Telescope (LST) of the planned Northern CTA observatory \citep{cta_lst_A,cta_lst_B} that is already operating, $95\%$ CL upper limits on the $\gamma$-ray flux towards LHAASO~J2108$+$5157 were recently derived at energies above $500$~GeV \citep{lstJ2108}. Figure~\ref{j2108_single} (left) demonstrates that even these flux upper limits can be used to put constraints on the hadronic best-fit models based on the LHAASO data. In particular, the flux upper limits derived from observations with LST-1 (shown with blue markers in Fig.~\ref{j2108_single} left) are in tension with the $68\%$~CL prediction of the $\gamma$-ray emission from the extrapolation of the best-fit Pevatron model to lower energies (shown with red shaded area and dashed line in Fig.~\ref{j2108_single} left). A fit of the available LHAASO data, constrained, in addition, to be compatible with the LST-1 flux upper limits, is shown by the blue line in Fig.~\ref{j2108_single} (left). The significance of the PTS for this combination of data is $S_\mathrm{PTS}$=$-$2.4, which can be interpreted as an indication that this source is not a Pevatron. However, additional data will be required for a decision with high statistical significance. As shown in Fig.~\ref{j2108_single} (left), the sensitivity of the full Northern CTA Observatory after complete construction and acquisition of $50$~h of data will allow for further constraining measurements, especially within the energy range from 1~TeV to 10~TeV. Figure~\ref{j2108_single} (left) also clearly demonstrates that very important constraints on the nature of LHAASO~J2108$+$5157 can be obtained from extensive observation with ASTRI Mini-Array. This array of 9 Cherenkov telescopes will be able to detect gamma-ray photons up to an energy of 300~TeV and will have an angular resolution $\sim 3'$ at the highest energies \citep{astri_sens}, 
much better than currently available. Operations will start, with an initial layout of 3 telescopes, in early 2024, and then in the final configuration by the end of 2025, early 2026 (S.~Scuderi, personal communication), with a delay of 4$-$6 months with respect to the timeline foreseen by \citet{ASTRI_Scuderi}. Although LHAASO~J2108$+$5157 is a rather faint source, being the search for Pevatrons one of ASTRI Mini-Array key science objectives \citep{ASTRI_DAi,ASTRI_Vercellone}, $\mathcal{O}(200)$ hours deep exposure of this promising Pevatron candidate can be foreseen.

As discussed in Sec.~\ref{data_analysis}, the attenuation of~$\gamma$-rays due to $\gamma\gamma$~absorption is neglected at energies below $100$-TeV, given that its effects are within our assumed minimum uncertainty of $\xi=20\%$. The spectral dataset for LHAASO~J2108$+$5157 contains three points at energies above $100$~TeV, with relative errors of $32\%$ at $126$~TeV, $144\%$ at $200$~TeV and $193\%$ at $500$~TeV. The previous conclusions regarding the PTS do not depend on the available spectral data above $100$~TeV. The significance of the PTS leads to $S_\mathrm{PTS}=-0.2$ when only LHAASO data at energies below $100$~TeV are fitted, and $S_\mathrm{PTS}=-1.2$ when the available flux upper limits from LST-1 are taken into account together with LHAASO E$<$100~TeV data.

In addition to LHAASO~J2108$+$5157, another $\gamma$-ray source, HESS~J1702$-$420A \citep{j1702}, without any clear counterpart below TeV energies, was recently discovered and is discussed as a Pevatron candidate. This new $\gamma$-ray source emerges as a sub-component of the previously known bright source HESS~J1702$-$420 \citep{j1702_2009} at energies above $\sim$30~TeV. A hadronic emission model and the association with a Pevatron are discussed in \cite{j1702} due to the presence of several molecular clouds detected along the line of sight and the $\gamma$-ray spectrum extending without indication of a clear spectral cutoff up to energies of at least $100$~TeV. The available spectral data are shown in Fig.~\ref{j2108_single} (right), for a minimum relative flux error of $\xi=20\%$. Within the hadronic emission model described in Sec.~\ref{data_analysis}, the best-fit index is found to be $\Gamma_\mathrm{P}=1.57\pm 0.18$, which is compatible with the result derived in \cite{j1702}. The lower limit on the hadrons energy cutoff is $436$~TeV (at $95\%$~CL) and the PTS is insignificant ($S_\mathrm{PTS}=1$). Similar to LHAASO~J2108$+$5157, it is therefore impossible to decide based on the PTS and the available data whether HESS~J1702$-$420 is associated with a Pevatron or not. Additionally, Fig.~\ref{j2108_single} (right) shows are the $\gamma$-ray flux sensitivities of two planned observatories in the Southern hemisphere. The figure suggests that the planned SWGO and the Southern CTA observatory will both allow probing the $\gamma$-ray flux predicted by the hadronic model that best fits the currently available data from HESS. The future SWGO observations of this region can provide very valuable E$>$100 TeV data, while CTA observations will allow probing the source spectrum down to sub-TeV energies with an unprecedented angular resolution. 

\section{Pevatron Test Statistic Profiles of Pevatron Candidate Sources}
\label{pts_profile}

The joint $\gamma$-ray data analyses of the Pevatron candidate sources presented in Sec.~\ref{sec_data_analyis_all} assume a Pevatron definition threshold of 1~PeV, as discussed around Eq.~\ref{eq_PTS} in Sec.~\ref{sec_spectral_sigs}.~With this assumption, the obtained values of $S_\mathrm{PTS}$ quantify the statistical significance and corresponding CL for a putative underlying hadron spectrum to extend beyond 1~PeV as a power-law, without indication of a cutoff. In other words, S$_\mathrm{PTS}$ quantifies whether the source can contribute to the CR spectrum above 1~PeV. However, taking into account the available joint $\gamma$-ray spectral data, none of the sources discussed in the previous section does robustly reach a 5$\sigma$ level for $S_\mathrm{PTS}$. 

The Pevatron threshold, i.e. the $E_\mathrm{cut,\,p}$ term in the numerator of Eq.~\ref{eq_PTS}, used for the calculation of $S_\mathrm{PTS}$, can be modified to quantify the contribution of accelerated particles to the CR spectrum above a given energy threshold. In other words, $S_\mathrm{PTS}$ can be profiled to extract up to which energy threshold a significant contribution to the CR spectrum can be expected from a given source. As discussed in Sec.~\ref{sec_intro}, there is evidence that the knee feature for proton and helium nuclei might be at energies around $700$~TeV, i.e. lower than $1$~PeV~\citep{knee_below_1pev}. In this case a Pevatron threshold of $\sim$300~TeV could be sufficient for a source to contribute to the proton knee, namely to the highest energy protons accelerated in the Galaxy.

\begin{figure*}
\centering
\includegraphics[width=17cm]{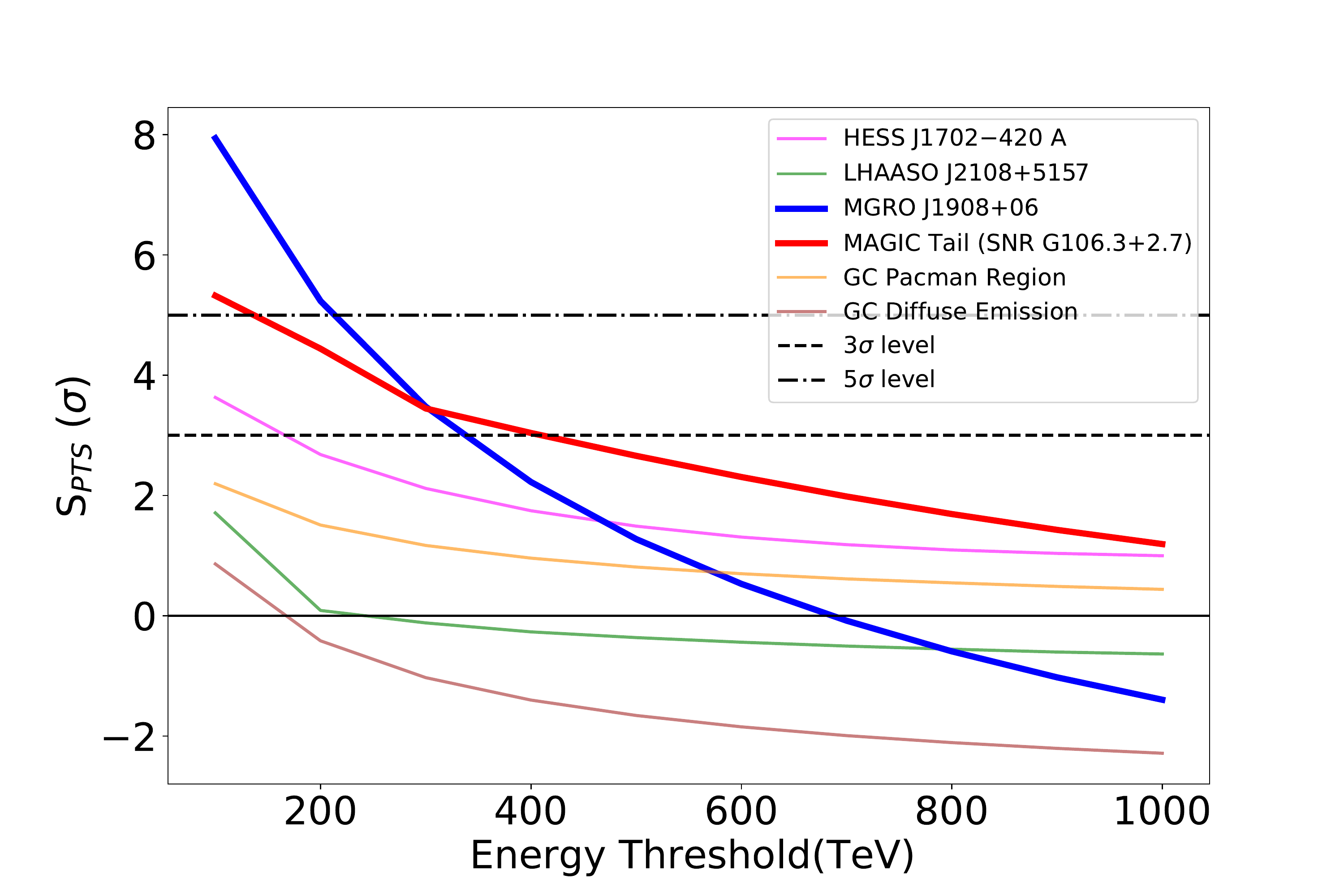}
\caption{Shown are the $S_\mathrm{PTS}$ profiles of the sources analyzed in this section.~$S_\mathrm{PTS}$ profiles of the emissions coming from the directions of the Magic Tail region (see Fig.~\ref{fig_boomerang}~right), MGRO~J1908$+$06 region (see Fig.~\ref{fig_j1908}~bottom left), HESS~J1702$-$420~A (see~Fig.~\ref{j2108_single}~right), LHAASO~J2108$+$5157 region (see~Fig.~\ref{j2108_single}~left),~GC~Pacman (see~Fig.~\ref{fig_gc}~bottom right) and Diffuse (see Fig.~\ref{fig_gc} bottom left) emission regions are shown with red, blue, magenta, green, orange and brown solid lines, respectively. The y-axis shows the corresponding $S_\mathrm{PTS}$ value obtained from Eq.~\ref{eq_PTS} using a given energy threshold. The~3$\sigma$ and~5$\sigma$ $S_\mathrm{PTS}$ levels are shown with dashed and dot-dashed black lines, respectively.} 
\label{profile_plot}
\end{figure*}

Figure \ref{profile_plot} shows the threshold energy dependent profile of $S_\mathrm{PTS}$ for the sources discussed in this work, which result in $|S_\mathrm{PTS}|<5$ for a Pevatron threshold of 1~PeV. The profiles were extracted using Eq.~\ref{eq_PTS} for a set of Pevatron energy thresholds between 100~TeV and 1~PeV with a step size of 100~TeV. It can be seen from this figure that for MGRO~J1908$+$06 (H+L, see~Fig.~\ref{fig_j1908} bottom left) and the tail region of SNR~G106.3$+$2.7 as seen by MAGIC (see~Fig.~\ref{fig_boomerang} right), a marginal significance level of 3$\sigma$ at energies around 350$-$400~TeV, and a robust 5$\sigma$ level at energies of 150$-$200~TeV, is reached. Assuming that the underlying emission mechanism is hadronic, these results provide marginal evidence that astrophysical objects responsible for the $\gamma$-ray emission seen from the direction of MGR0 J1908$+$06 and the tail region of the SNR~G106.3+2.7 seen in MAGIC data analysis can contribute to the knee of proton (and helium) spectra when the knee feature for these light elements is at energies around $700$~TeV. Similarly, for the Pevatron candidate source HESS~J1702$-$420A, a marginal 3$\sigma$ level is reached for threshold energies around 200~TeV. Eventually, the S$_\mathrm{PTS}$ profile for the GC Pacman region does not reach a $3\sigma$ level for energies above $100$~TeV and is therefore less promising.~However, there are currently no UHE data available for the GC Pacman reach and HESS~J1702$-$420A and the spectral data which will be acquired with future observations by SWGO have key importance and can potentially increase the achieved $S_\mathrm{PTS}$ levels. 

\section{Conclusion}
\label{sec_conclude}

In this work, a Pevatron is defined to be a source of CRs at energies around the knee of the CR spectrum. Based on this definition, the PTS is shown to be a unified metric for the confirmation and exclusion of an association between $\gamma$-ray sources and Pevatrons which exhibits clear advantages over other currently employed methods, and offers a new approach for the robust detection of Pevatrons. As demonstrated in this paper for multiple Galactic $\gamma$-ray sources, the method is simple to apply in practice, especially for isolated sources and resolved source components. With a statistical significance of more than $5\sigma$, it is excluded that the two shell type SNRs RX~J1713.7$-$3946 and Vela~Jr. are Pevatrons that can contribute to the knee feature seen at $\sim$3~PeV energies. Similarly, the Pevatron hypothesis for the Galactic central source HESS~J1745$-$290 can also be excluded with a significance level of more than $4\sigma$. The importance of using high angular resolution observations to resolve source confusion when searching for Pevatrons is demonstrated with the PTS analysis of the $\gamma$-ray emission region encompassing the SNR~G106.3$+$2.7 and the Boomerang nebula. In this region source confusion is problematic. The PTS analysis results for the case when the region is considered as a unique source and when it is resolved in two sources are compared to each other, leading respectively to $S_\mathrm{PTS}$=-5.2$\sigma$ and 1.2$\sigma$, while the corresponding 95$\%$ C.L. lower limits on the proton cutoff are found to be $\sim$240~TeV and $\sim$820~TeV, respectively. This demonstrates clearly that source confusion can lead to misleading total $\gamma$-ray spectra, possibly obscuring Pevatron signatures, and implies the critical importance of high angular resolution observations for Pevatron searches, especially at energies above 10~TeV.  No statistically significant conclusion can be drawn for the unidentified sources LHAASO~J2108+5157, HESS~J1702-420A and MGRO~J1908+06. However, it is argued that data from future observatories, like the CTA, ASTRI Mini-array, and SWGO will help to decide whether these sources are Pevatrons. With currently available data, we tried to determine up to what energies these sources can contribute to the CR spectrum. Assuming a purely hadronic origin of the $\gamma$-ray emission, we found that the parent proton spectra of MGRO~J1908$+$06 and the tail region of SNR~G106.3$+$2.7 can reach marginal PTS levels of $3\sigma$ at energies around 350$-$400~TeV, and even $5\sigma$ at energies around 200~TeV. This result is a strong indication for these two sources being proton and helium Pevatrons, and likely contribute to the knee of the proton and He spectra around 700~TeV observed at Earth.

\section*{Acknowledgements}
E.O.A.~acknowledges financial support by TÜBİTAK Research Institute for Fundamental Sciences.\\ G.S.~acknowledges financial support by the German Ministry for Education and Research (BMBF).\\ S.C.~acknowledges financial support from the Polish National Science Centre, grant DEC-2017/27/B/ST9/02272.\\
E. A.~acknowledges financial support by INAF under grant INAF-MAINSTREAM 2018 and PRIN-INAF 2019.\\ 
This research has made use of the CTA instrument response functions provided by the CTA Consortium and Observatory, see \url{https://www.cta-observatory.org/science/cta-performance/} version prod5 v0.1 \cite{prod5} for more details.\\
This research has made use of the ASTRI Mini-Array sensitivity curve provided by the ASTRI Project \citep{astri_irfs}, see \cite{astri_sens} for more details. We are grateful to Saverio Lombardi and Stefano Vercellone for their comments and indications in relation to the ASTRI Mini-Array performance, and to Salvatore Scuderi for updates on the ASTRI Mini-Array timeline.
\\
We express our sincere gratitude to Heide Costantini, Kathrin Egberts, and Ulisses Barres de Almeida for their useful contributions and constructive feedback, which greatly enhanced the quality of the paper.\\
\vspace{5mm}\\
\textbf{Facilities : CTA, Fermi, HAWC, HESS, LHAASO, MAGIC, SWGO, VERITAS, ASTRI Mini-Array}
\section*{DATA AVAILABILITY}
The data that support the findings of this study are openly available and taken from the respective publications which are explicitly mentioned in the figures and text.
\section*{SOFTWARE}
The calculations are performed with {\tt ecpli} python package \citep{ecpli}, which uses {\tt naima} \citep{naima} and {\tt gammapy} \citep{gammapy_v018} python packages.\\





\bibliographystyle{mnras}
\bibliography{main_pts_paper} 



\bsp	
\label{lastpage}
\end{document}